# Flexible Perovskite/Silicon Monolithic Tandem Solar Cells Approaching 30% Efficiency

Yinqing Sun[1]†, Faming Li[1]†, Hao Zhang[1]†, Wenzhu Liu[2]†, Zenghui Wang[3], Lin Mao[1], Qian Li[1], Youlin He[1], Tian Yang[1], Xianggang Sun[1], Yicheng Qian[1], Yinyi Ma[1], Liping Zhang[2], Junlin Du[2], Jianhua Shi[2], Guangyuan Wang[2], Anjun Han[2], Na Wang[2], Fanying Meng[2], Zhengxin Liu[2]*, Mingzhen Liu[1,4]*

[1]School of Materials and Energy, University of Electronic Science and Technology of China, Chengdu 611731, P.R. China

[2]Research Center for New Energy Technology, Shanghai Institute of Microsystem and Information Technology, Chinese Academy of Sciences, Shanghai, P.R. China

[3]Institute of Fundamental and Frontier Sciences, University of Electronic Science and Technology of China, Chengdu 610054, P.R. China

[4]State Key Laboratory Electronic Thin Film and Integrated Devices, University of Electronic Science and Technology of China, Chengdu 611731, P.R. China

†These authors contributed equally to this work.

*Corresponding authors: z.x.liu@mail.sim.ac.cn; mingzhen.liu@uestc.edu.cn

## Abstract

Thanks to their excellent properties of low cost, lightweight, portability, and conformity, flexible perovskite-based tandem solar cells show great potentials for energy harvesting applications, with flexible perovskite/c-silicon tandem solar cells particularly promising for achieving high efficiency. However, performance of flexible perovskite/c-silicon monolithic tandem solar cells still greatly lags, due to challenges in simultaneously achieving both efficient photocarrier transport and reliable mitigation of residual stress. Here, we reveal the critical role of perovskite phase homogeneity, for achieving high-efficient and mechanical-stable flexible perovskite/c-silicon heterojunction monolithic tandem solar cells (PSTs) with textured surface. Through ensuring high phase homogeneity, which promotes charge transfer across all facets of the pyramid on the textured substrates and releases the residual stress at the perovskite/c-silicon interface, we demonstrate flexible PSTs with a bending curvature of 0.44 $cm^{-1}$, and a certified power conversion efficiency of 29.88% (1.04 $cm^2$ aperture area), surpassing all other types of flexible perovskite-based photovoltaic devices. Our results can lead to broad applications and commercialization of flexible perovskite/c-silicon tandem photovoltaics.

## Introduction

Metal halide perovskites have been widely used for ultra-thin, flexible solar cells, thanks to their high absorption coefficient and low Young's modulus[1-3], offering great promises for wearable electronic devices, building-integrated photovoltaics (BIPV), portable energy systems, and aerospace technologies[4,5]. Both single-junction flexible perovskite solar cells (f-PSCs)[6-8] and flexible perovskite-based tandem devices, such as perovskite/perovskite[9], perovskite/Cu(In,Ga)$Se_2$ (CIGS)[10,11], and perovskite/organic photovoltaics (OPV)[12], have been actively and broadly developed, demonstrating their excellent potential in enabling next-generation high-efficiency flexible photovoltaics (PV) devices.

Among them, tandem structure based on wide-bandgap perovskite and narrow-bandgap c-silicon (c-Si) heterojunction solar cell, with its promising power conversion efficiency (PCE) and great stability, has demonstrated great promises to become a mainstream technology in the PV market[13-16]. Especially, using c-Si wafers with reduced thickness (to less than 100 micrometers)[17,18] and blunted edges[19], superb foldability and high response to infrared wavelengths can be simultaneously achieved, offering excellent potential of designing efficient flexible perovskite/c-Si tandem solar cells.

However, such great opportunities have remained largely unexplored to date[20], significantly limiting the improvement of PCE in perovskite-based flexible PVs. The primary challenge lies in simultaneously achieving both efficient longitudinal transport of photocarriers[21,22], and reliable mitigation of residual stress which could induce delaminating and fracturing[23,24], crucial for flexible tandem devices. This is mainly caused by inadequate reaction[25,26] and nonideal cation distribution[27,28], which is further complicated by the use of perovskite/c-Si tandem solar cells with textured surface for enhanced light-trapping.

Here we tackle these challenges to demonstrate highly efficient and reliable flexible perovskite/c-Si heterojunction monolithic tandem solar cells (PSTs). We reveal the critical role of phase homogeneity of perovskite films in flexible PSTs with textured substrates. We find that highly homogeneous phase distribution of perovskite films not only effectively promote the charge transport and extraction across all facets of the pyramid on the textured substrates, but also efficiently release the tensile stress between the perovskite layer and textured c-Si bottom cell during bending. The optimized flexible PSTs achieve a champion certified PCE of 29.88%, along with superior mechanical endurance of maintaining initial efficiency after 2000 bending cycles. Furthermore, we demonstrate that the mechanical endurance of flexible PSTs can be extended to the fracture limit of Si substrate, only when highly-phase-homogeneous perovskite is achieved. Our findings pave the pathway towards high performance flexible perovskite-based solar cells for high-power-to-weight-ratio, high-conformability, high-integrability power applications.

**Fabricating flexible perovskite/c-Si heterojunction monolithic tandem solar cells**

We use a flexible c-Si heterojunction solar cell with a thickness of ~70 micrometers and blunted edges as the bottom cell and fabricate perovskite layer on top to form tandem solar cells (*Fig. 1a-b*), which contains the indium tin oxides (ITO) recombination layer, the hybrid hole-transporting layer ([2-(9H-carbazol-9-yl)ethyl]phosphonic acid (2PACz)/nickel oxide ($NiO_x$) layer composite), the perovskite layer, and other functional layers.

We start with vapor-phase growth by evaporating Cs source to provide an ultra-uniform inorganic framework, followed by spinning organic cation solution (*Fig. 1c*). Particularly, we focus on the distribution of organic cations, which determines the phase homogeneity throughout the perovskite films and thus the band alignment, carrier transport, and residual stress. Homogeneity of perovskite phase distribution is critical for perovskite solar cells on flat substrate[28], which is both more important and challenging on textured substrates. Specifically, we compare the effects of mixed (formamidinium, FA /methylammonium, MA, control film) and single (FA, target film) cations during the liquid-phase fabrication, aiming to form homogeneous perovskite phases on the textured substrates.

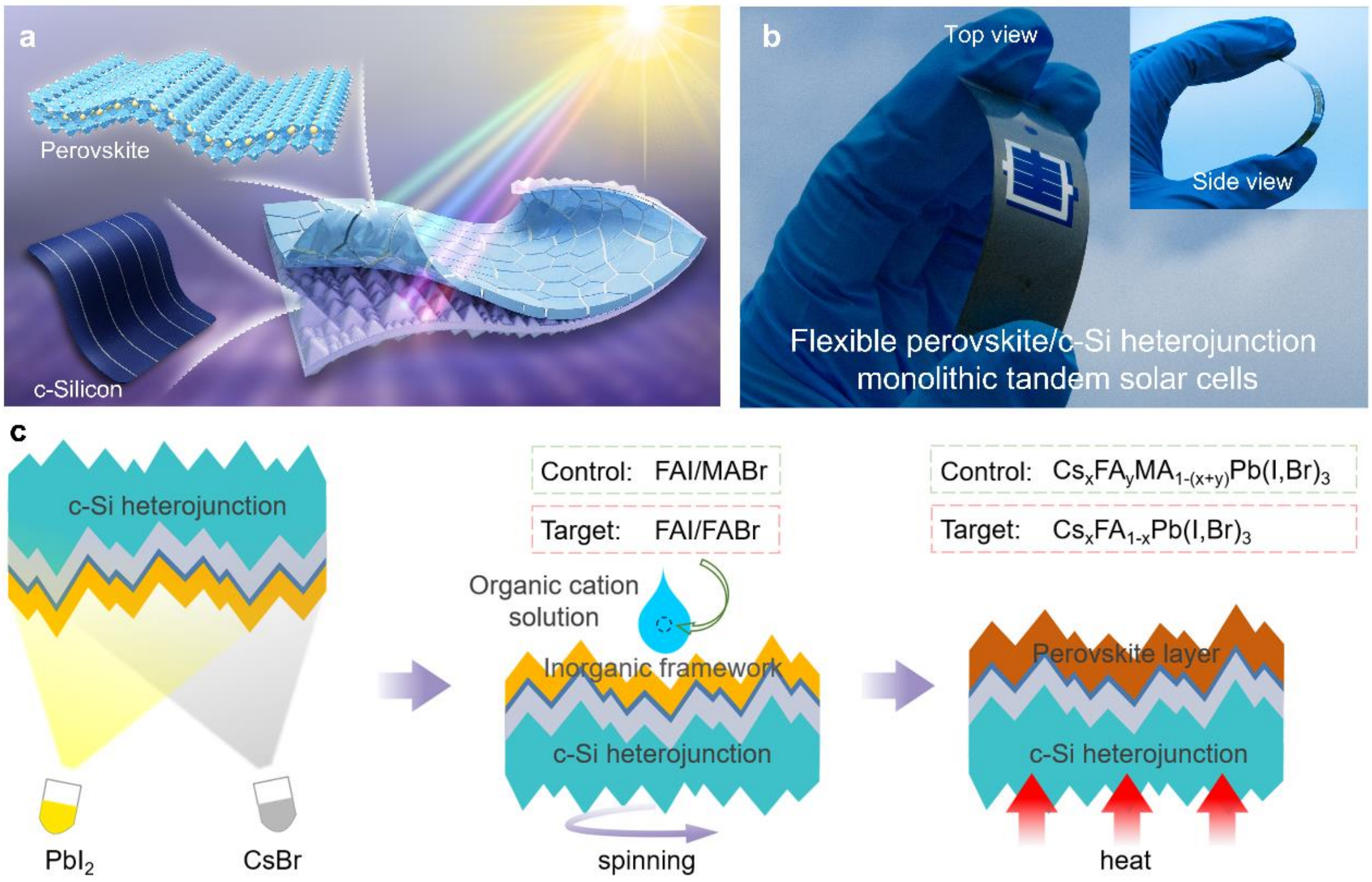

**Figure 1**. **Schematics of flexible PSTs. a,** Structural schematic of flexible PSTs. **b,** Digital photo of the flexible tandem solar cells. **c,** Schematic illustration of the synthesis process of perovskite on flexible textured c-Si heterojunction bottom cells.

## Characterizing and understanding perovskite phase homogeneity

To directly observe the phase distribution on textured substrates from a microscopic perspective, we first use cross-sectional transmission electron microscope (TEM) to examine the lattice homogeneity for top, middle, and bottom parts of perovskite films (*Figs. 2a-b*). In stark contrast to the control film (*Fig. 2a*), which exhibits varying lattice spacings $d$ (ranging from 3.45 Å to 2.79 Å), the target film (*Fig. 2b*) exhibits high identical lattice spacing throughout (3.42 Å). The consistent pattern of lattice spacing along the depth direction is also observed at the crest and trough area of the pyramids (*Supplementary Fig.1*). This observation suggests that the use of single organic cation can produce homogeneous perovskite lattice, while the mixed organic cations with different ionic radii give rise to varied lattice spacing, with organic cations more accumulated on the top and lacking at the bottom. Notably, perovskite on textured substrates is much more sensitive and selective to the choice of cations than those on flat substrates (*Supplementary Fig. 2*). This underscores that achieving phase homogeneity is fundamentally more challenging on textured substrates. Such contrast in homogeneity between control and target samples is further evidenced by photoluminescence (PL): the control film exhibits different PL peak positions (by ~14 nm) for top and bottom surface of the film, while the target film shows much more consistent emission peaks (*Fig. 2c and Supplementary Fig. 3*).

The phase inhomogeneity observed in the control film can be caused by the difference in perovskite formation energy involving $FA^+$ and $MA^+$ cations. $MA^+$ cations give a lower formation energy ( $\triangle E$=-1.67 eV) for the perovskite crystallization (*Fig. 2d and Supplementary Fig.4*), and tends to form perovskite crystals at the surface before diffusing into the inorganic framework. This is also confirmed by focused ion beam-time-of-flight secondary-ion mass spectroscopy (FIB-ToF-SIMS), showing accumulation of $MA^+$ and $FA^+$ ions on the control film top surface (*Supplementary Figs. 5 and 6*). In contrast, $FA^+$ cations in the target film can diffuse more uniformly throughout the inorganic framework thanks to its higher formation energy ($\triangle E$=0.23 eV), resulting in retarded crystallization for the target film. Such difference is further evidenced by real-time in-situ confocal laser scanning microscopy (CLSM), with the target film showing a delayed rise of PL lifetime (indicating crystallization) of 25 s compared to the control film (13 s) (*Fig. 2e and Supplementary Fig. 7*). Therefore, using single organic cation with higher formation energy is desirable for achieving phase-homogeneous perovskite films.

To evaluate the enhancement of charge transfer by the phase homogeneity, we compare the quenching effect by depositing a hole transport layer (HTL) or electron transport layer (ETL) on half of each perovskite film. CLSM mapping shows more effective carrier extraction (larger color contrast in *Fig. 2f and Supplementary Fig.8*) by either the HTL or ETL in the target film. This is further confirmed by charge selectivity measurements using conductive atom force microscope (c-AFM), with larger current (*Fig. 2g*) found for the target film in both positive and negative bias.

Such superior charge transfer is attributed to the highly-aligned band structure thanks to the phase homogeneity. Ultraviolet photoelectron spectroscopy (UPS) results of top and bottom surface of perovskite film show much better band alignment in the target film, while the control film shows skewed band structure (*Fig. 2h and Supplementary Fig. 9*), in which tilted bands hinders charge transfer along the depth direction[29,30].

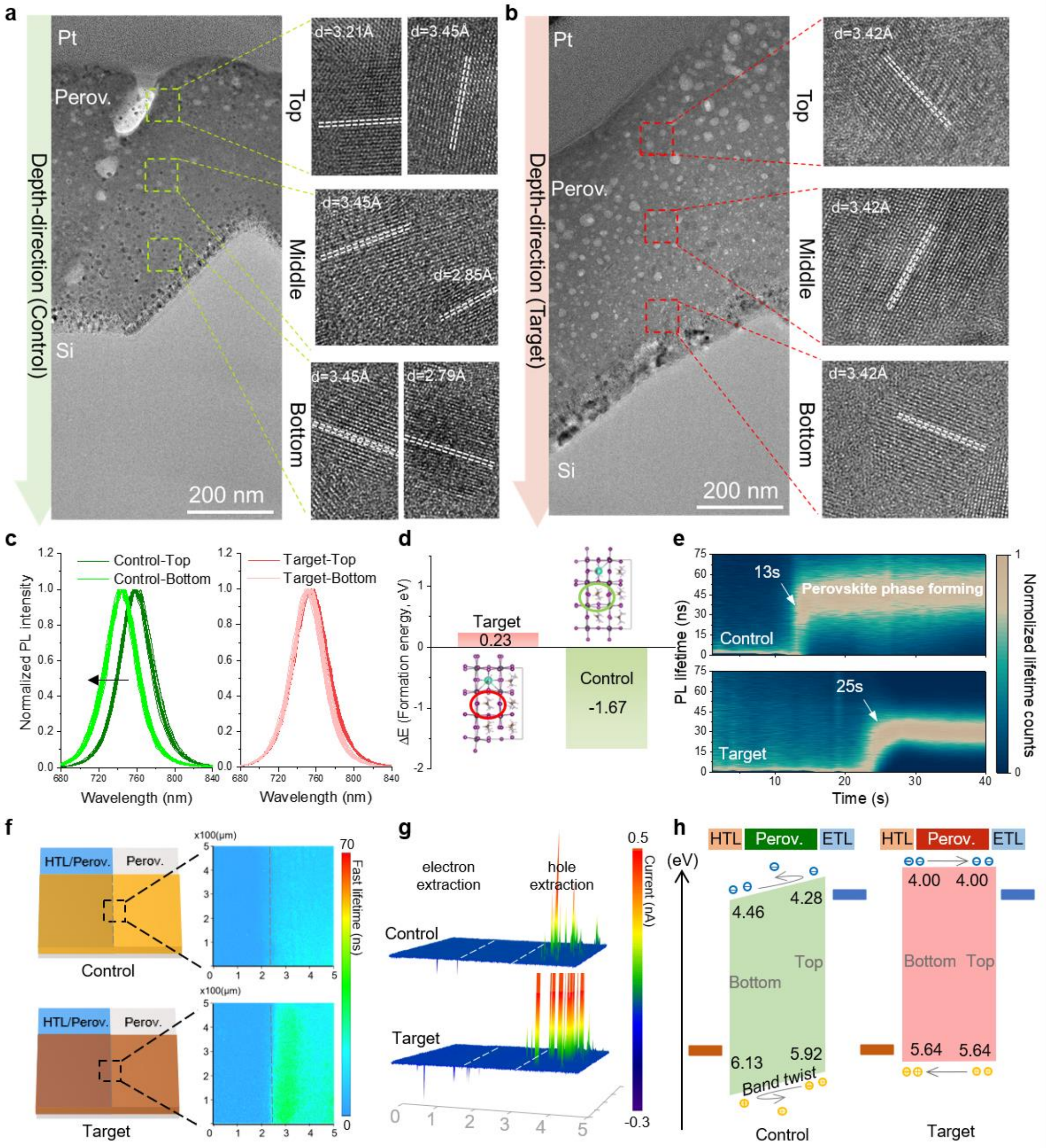

Figure 2. **Investigation of perovskite phase homogeneity on textured silicon substrates and its impacts on carrier transfer. a,b,** High-resolution TEM images collected from the corresponding squares of (**a**) control and (**b**) target perovskite (Perov.) film respectively. **c,** PL emissions of 20 points from top and bottom surface respectively. **d,** Calculated formation energies for perovskites. **e,** Real-time in-situ CLSM. **f,** Direct observation of PL quenching of control and target films by the extraction effect of HTL. **g,** 3D c-AFM images ($5 \times 5$ μm, electrical current) of the samples with a structure of c-Si/ITO/$NiO_x$/2PACz/Perov./$C_{60}$. **h,** Schematics for the band alignment of perovskite film.

## Mechanical endurance of perovskite films on textured substrates

The lattice and phase homogeneity along the depth of perovskite films regulates the residual stress, and thus adhesion at textured interfaces, which is particularly crucial for flexible tandem solar cells during bending[31,32]. We correlate the residual stress with the perovskite lattice using grazing-incidence X-ray diffraction (GIXRD). The lattice spacing of target film exhibits minimal variation with Ψ angles (*Supplementary Fig.10*), with a residual stress as low as 23.3 MPa. In contrast, the control film shows a significant increase in lattice spacing, with a residual stress as high as 48.7 MPa, more than doubling that of the target film.

To visualize the resistance to bending as a result of stress release, we compare the thin-film topology and cross-sectional structure of the two perovskite films before and after bending. The top-view scanning electron microscope (SEM) images (*Figs. 3a-d*) show that the control film exhibits multiple distinct fracture lines (indicated by the arrows) upon bending, while such defects are not observed in the target film. To further analyze the effects of fracture lines on the textured substrates, we use focused ion beam (FIB)-cutting to examine the cross section of the perovskite layer on textured c-Si bottom cell (*Figs. 3e-h*). As seen in *Fig. 3g* and *3h*, after 2000 cycles the control film suffers cracks and delamination from the textured substrates, while the target film remains largely intact. Notably, the appearance of fracture lines and interfacial delamination is spatially highly correlated. We also perform in-situ TEM test during bending using a specialized inset that allows application of force in-situ. We observe that the stress lines tend to aggregate on the troughs of the pyramids (*Supplementary Fig.11*), indicating that the interfacial delamination is highly concentrated spatially (*Fig. 3g*). This further indicates the importance of the perovskite phase homogeneity along the depth from crest to trough of such pyramids on the textured substrates, which can effectively facilitate the release of the residual stress and thus significantly enhances endurance during bending.

In addition to the crack/delamination, another considerable difference observed between the perovskite films is the quality of crystallization. The target film exhibits much higher level of crystallization with well-defined grain boundaries, in contrast to the control film. This again showcases the benefit of the retarded crystallization, which enhances the charge transport in the resulting flexible device by reducing the density of the grain boundaries (which are likely to impair charge transfer under mechanical deformation).

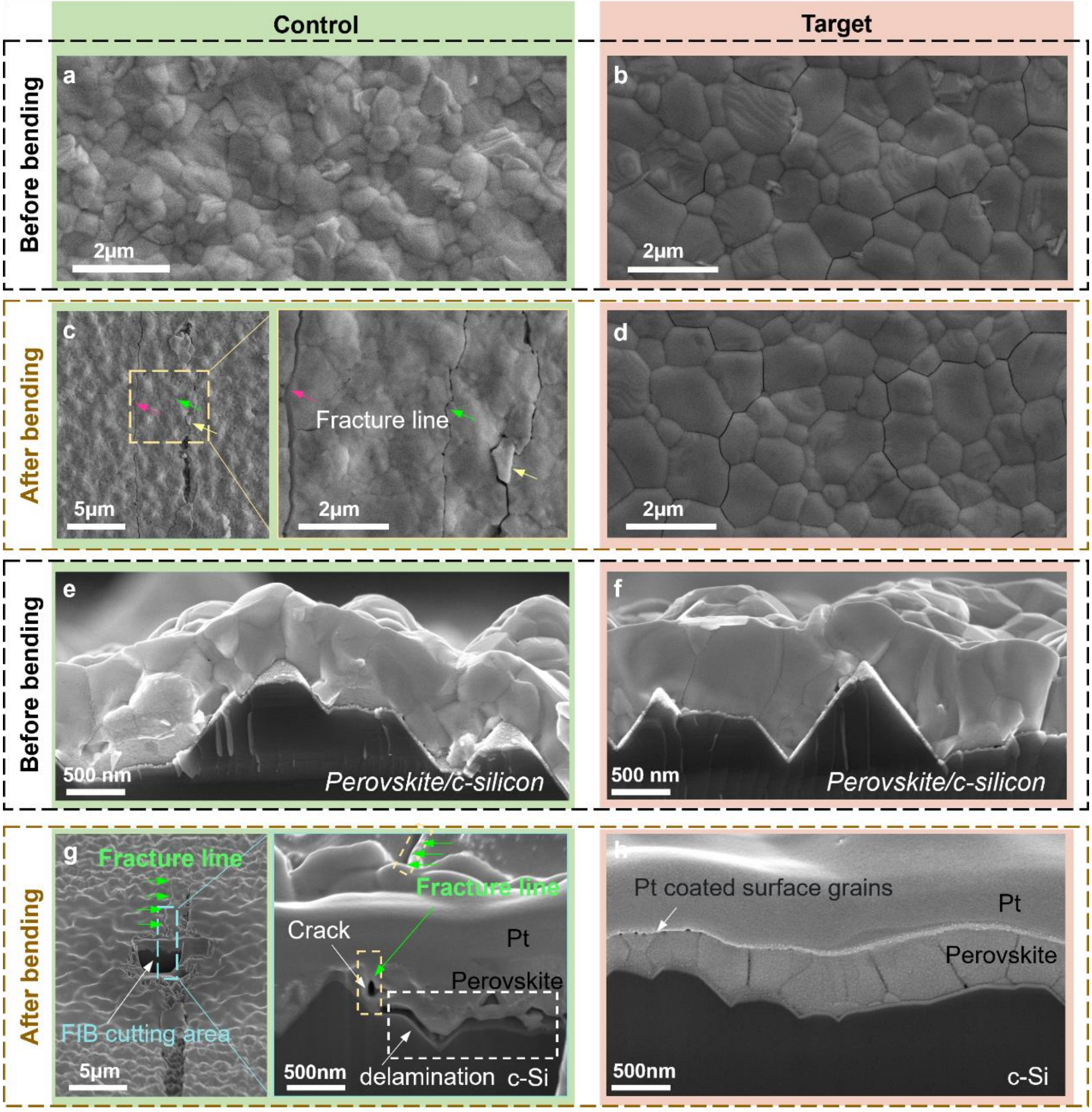

Figure 3. **Film morphology of perovskite films on textured substrates before/after mechanical durability tests. a-d,** Top-view SEM images of control and target perovskite on flexible textured silicon substrates (**a, b**) before and (**c, d**) after 2000 bending cycles, respectively. **e,f,** Cross-section SEM images of (**e**) control and (**f**) target perovskite on flexible textured silicon substrate before 2000 bending cycles. **g,h,** Cross-section FIB-SEM images of (**g**) control and (**h**) target perovskites on flexible textured silicon substrate after 2000 bending cycles .

### Performance and flexibility of tandem solar cells

We also evaluate the PV performance of the flexible PSTs, prepared with device architecture of Ag/ITO/a-Si(p)/a-Si(i)/n-Si/a-Si(i)/a-Si(n)/ITO/HTL/Perovskite/ETL/$SnO_2$/Indium Zinc Oxide (IZO)/Ag/LiF (*Fig. 4a*). The devices using the target films show superior performance compared to the control ones. The optimized device exhibits a short-circuit current density ($J_{sc}$) of 19.51 mA/cm$^2$, a fill factor of 0.809, and an open-circuit voltage ($V_{oc}$) of 1.89 V, yielding a champion PCE of 29.83% with an active area of 1.04 cm$^2$ (surpassing the control device with 27.45% PCE) measured under simulated sunlight (*Fig. 4b, Supplementary Figs. 14 and 15*). The integrated $J_{SC}$ from external quantum efficiency (EQE) response of perovskite and silicon subcells are 19.10 and 19.21 mA/cm$^2$, respectively *(Fig. 4d)*. We also determine the stabilized power output of this champion flexible PSTs by measuring the current at the maximum power point (MPP) voltage over 80 s with a stabilized PCE of 29.20% (*Fig. 4c*). To further validate our results, we send our devices to National Institute of Measurement and Testing Technology (NIMTT, Chengdu, China) and the best device achieves a certified PCE of 29.88% (*Fig. 4e, Supplementary Figs. 16 and 17*). We further evaluate the stability of the devices using the procedure specified in the ISOS-L-1 protocols[33], and age them under one-sun LED illumination (25 °C). After 1000 hours of continuous illumination at the MPP in a nitrogen atmosphere, the unencapsulated target device remains stable and constant, while the control device drops to 92% of its initial efficiency (*Fig. 4f*). These results again demonstrate the superiority of perovskite films with uniform phase distribution, even at the device level.

To demonstrate the flexibility of PSTs from a practical level, we develop a three-point bending test and observe that our flexible tandem devices can endure significant deformation (*Fig. 4g and Supplementary Fig. 18*). We further investigate the endurance to repeated bending of flexible tandem devices by assessing the PCE variation with different bending radii and bending cycles (*Supplementary Fig.19*). As the bending radius decreases to 2.25 cm, corresponding to a curvature of 0.44 $cm^{-1}$, the target device is able to maintain its initial efficiency, while the control device shows significantly falls to 52% of its initial efficiency (*Fig. 4h and Supplementary Fig.20*). We also study device durability under 3.2 cm radius bending. The target devices maintain its 100% initial value after full 2000 cycles, while the PCE of control devices drops by more than 60% after just 1000 cycles (*Fig. 4i*). This suggests that the mechanical endurance of flexible PSTs can be extended to the fracture limit of Si substrate. The improved mechanical durability of the target device is consistent with the enhanced interfacial adhesion at the perovskite/textured silicon interface thanks to the uniform phase distribution of perovskite films by releasing residue stress. Overall, our findings provide an in-depth investigation of phase homogeneity in flexible perovskite/c-Si monolithic tandem solar cells and suggest a promising path for further promoting the application of flexible photovoltaic technology.

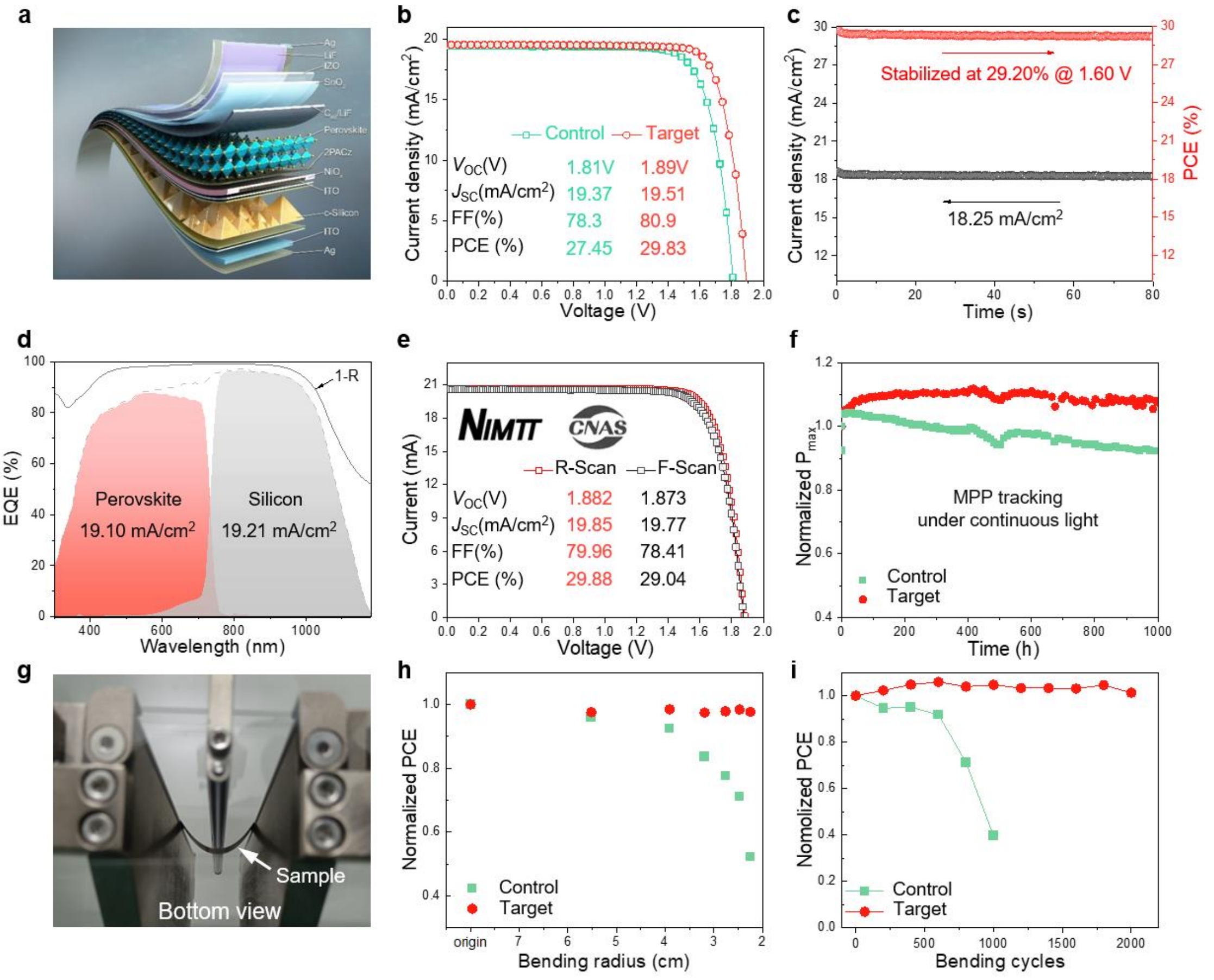

Figure 4. **Device performance of flexible PSTs**. **a,** An artist's schematic illustration of the flexible tandem device. **b,** The J-V curves and PV parameters (inset) of control and target devices, the device performance of single junction silicon solar cells and perovskite solar cells are shown in *Supplementary Figs. 12 and 13*. **c,** Steady-state PCE and current density of the champion target devices. **d,** EQE plots for the target device. **e,** Independent performance certification from National Institute of Measurement and Testing Technology. **f,** Normalized PCE evolution of flexible PSTs after continuous exposure to one-sun light in $N_2$ environment for 1000 hours. **g,** Diagram of the Discovery dynamic mechanical analyzer (DMA) 850 for three-point bending test. **h,i,** Bending tests of flexible tandem devices with (**h**) different bending radii and (**i**) different bending cycle numbers at bending radius of 3.2 cm. Additional data is presented in *Supplementary Fig. 21 and Supplementary Table 1*.

## Methods

### Materials

All reagents and solvents are used without purification. Lead iodide ($PbI_2$, 99.99%), Cesium bromide (CsBr), and methylammonium chloride (MACl) are purchased from Xi'an Polymer Light Technology Corp. Formamidine iodide (FAI), methylammonium bromide (MABr) and formamidine bromide (FABr) are purchased from Greatcell solar materials. Fullerene ($C_{60}$), lithium fluoride (LiF) and ethanol are purchased from Sigma-Aldrich. The ceramic 2-inch $NiO_x$, indium-doped tin oxide (ITO, In:Sn=95:5 wt.%) and indium zinc oxide (IZO, In:Zn=90:10 wt.%) targets are purchased from Kairui New Materials co. ltd.

### Device fabrication

*Flexible Silicon Heterojunction Solar Cell.* As reported by our previous work[19], the n-type Czochralski (CZ) c-Si wafers with 70 μm thickness are used as the bottom subcells. The Czochralski n-type c-Si wafers initially hold a thickness of 160 μm. To remove the saw damage, they are treated in a 20.0-vol% alkaline water solution at 80 °C, with varying treatment times to achieve various wafer thicknesses. Subsequently, the wafers are immersed in a 2.1 vol% alkaline solution at 80 °C for 10 minutes to form a microscale pyramid on both sides of the wafer surface. Approximately 2 mm wide edge region of these 70 μm textured wafers are passivated in in a HF: $HNO_3$ (10: 90 vol%) solution for 90 seconds at 25 °C to achieve better flexibility. All wafers are cleaned through the standard RCA cleaning process and subsequently soaked in a 2.0% hydrofluoric acid aqueous solution to eliminate organic matter and metal ions and etch surface oxides, respectively. Then, 5 nm i-a-Si: H and 15 nm p-a-Si:H, as well as 4 nm i-a-Si:H and 6 nm n-a-Si:H, are deposited on the back and front, respectively, at a process temperature of 200 ℃ in two RF plasma-enhanced chemical vapor deposition (PECVD). Indium tin oxide (ITO) is sputtered on the surface of N-type layer and P-type layer respectively. The thickness of ITO on the surface of N-type layer is 20 nm and the series resistance is 300 Ω per square; the thickness of ITO on the surface of P-type layer is 80 nm and the series resistance is 50 Ω per square. Following that, 200 nm Ag is evaporated as the back electrode.

*Flexible Tandem Solar Cell.* The flexible silicon bottom cells were cut into 2×2 $cm^2$ or 2×6 $cm^2$ squares by laser and heated for 30 minutes. Then, at 0.37 Pa, 25 °C and 90 W power, the hole transfer $NiO_x$ layer was RF sputtered on the surface of the textured silicon for 10 minutes. The final thickness of $NiO_x$ layer is 30 nm. The 50 μL of 2PACz (1 mg/mL in anhydrous EtOH) solution was spin-coated on the $NiO_x$ layer at 3000 rpm for 30 s and then annealed at 100 °C for 10 min. The perovskite film is deposited using a vapor/solution hybrid two-step method, with $PbI_2$ and CsBr co-evaporated at a rate of 10:1 to a thickness of 350 nm. Then, a mixture of FAI/FABr(MABr) is dissolved in ethanol at concentrations of 61/20(15) mg/mL. 5 mg/ml MACl is added to achieve better crystallinity. The solution is spin-coated at 4000 rpm for 30 s in ambient air with 5% RH, followed by annealing step at 150 °C for 30 min in ambient air with 75% RH. The final composition of the control perovskite is $Cs_xFA_yMA_{1-(x+y)}Pb(I,Br)_3$ ($0<x+y<1$) and that of the target perovskite is $Cs_xFA_{1-x}Pb(I,Br)_3$ ($0<x<1$). Following the thermal evaporation of a 20 nm thick $C_{60}$ film as the electron transport layer (ETL), a 10 nm thick layer of $SnO_2$ is subsequently deposited via atomic layer deposition to serve as a buffer layer, which aims to prevent sputtering damage. Thereafter, a 120 nm indium zinc oxide (IZO) film is deposited by direct current (DC) sputtering at 25 °C. A 600 nm thick Ag grid is thermally evaporated onto the front surface using a shadow mask, and subsequently, a 200 nm thick Ag grid is also evaporated onto the back surface. Eventually, antireflection layer of LiF of 100 nm thickness is evaporated.

**Material characterizations**

*Revealing the "Bottom" of perovskite film on the textured substrates*[34,35]. A solution of PTAA at a concentration of 10 mg/ml in chlorobenzene is spin-coated onto a silicon substrate, and the deposition of the perovskite is carried out as previously described. Following 2 hours immersion in chlorobenzene, the bottom surface is delaminated from the textured silicon substrates.

*Deriving the bandgap values from different transport layer side*: First, we place the sample in the CLSM setup, and select a test area of 200 × 200 μm. Then, we pick 10 sample points in each of the two orthogonal directions, and all the 20 spectra are averaged. Finally, the bandgap of perovskite is deduced using the position of corresponding emission peaks.

*Focused ion beam-transmission electron microscope (FIB-TEM).* The samples are prepared with c-Si/$NiO_x$+2PACz/perovskite stacks. The surface of the sample is protected by depositing Pt and carbon layer before thinning with focused ion beam (FIB) equipment (Thermo Fisher Helios 5 CX). Pt is deposited using 5 kV, 0.34 nA electron beams and 80 pA ion beams. After Pt deposition, excavation begins at 30 kV, 2.5 nA and thinned at 30 kV, 0.43 nA. After the extraction of the FIB-sample, the thinning begins from a low voltage with 0.79 nA, followed by 0.23 nA, and finally decreases to 80 pA. Then the TEM test is carried out in ThermoFisher Talos F200S.

*Focus ion beam-time of flight secondary ion mass spectrometry (FIB-ToF-SIMS).* The test is conducted using a PHI nanoToF instrument equipped with a GA source and FIB accessories. The FIB processing is performed prior to the ToF-SIMS analysis, which is carried out directly afterward. During the analysis, the ion beam scans across the sample line by line, acquiring a mass spectrum at each pixel. To minimize interference and prevent unnecessary signals, the ion beam is blanked when transitioning between scan rows.

*Steady-state photoluminescence (PL).* The PL emission test is performed by PicoQuant FluoTime 300 with a 405 nm laser (LDH-P-C-405, Pico Quant GmbH).

*Conductive Atomic Force Microscopy (c-AFM).* We deposit a 10 nm silver layer atop the c-Si/ITO/$NiO_x$/2PACz/perovskite/$C_{60}$ stack (on textured silicon) to improve conductivity during c-AFM testing. The c-AFM test is performed with a Bruker Dimension Icon instrument. The tip used is model TESPA-V2. The bias is applied to the sample.

*Confocal laser scanning microscopy (CLSM).* The CLSM measurement is conducted by TCSPC module (MT200, Pico Quant). And the in-situ CLSM testing is carried out via continuously grabbing frames at the rate of 1 frame per second for a total time scale of 40 seconds immediately after dripping the organic cation solution. The instrument focal length is pre-aligned with a pre-prepared perovskite film before testing.

*Ultraviolet photoelectron spectroscopy (UPS).* The UPS measurement is performed with a Thermo Scientific Escalab 250Xi instrument utilizing a He discharge lamp with a photon energy of 21.22 eV.

*Grazing-incidence X-ray Diffraction (GIXRD).* GIXRD test is conducted by Rigaku Smartlab 3kw using Cu Kα ($\lambda = 1.5406$ Å) radiation. And the data are detected at grazing incident angle of 0.5 ° with various Ψ angle of 0, 25, 35, 45, 55 °. The scan rate is 2.5 °/min. The samples are prepared on flat substrates, and the results show clear difference between control and target samples. It is important to note that such difference would be more significant for samples prepared on textured substrates, evidenced by data such as shown in Supplementary Fig. 2. Therefore, such non-trivial difference observed using flat-substrate samples fully supports the conclusion for textured-substrate samples.

*Scanning electron microscope (SEM).* Field-emission SEM (Thermo Scientific Helios 5 CX) are conducted to observe the film surface and cross-section morphologies before/after bending tests.

*In-situ transmission electron microscope (in-situ TEM).* The in-situ bending test is carried out on the Fei TECNAI F30 TEM system using the transmission in-situ sample rod of picofemto model. A c-Si/ITO/HTL/perovskite half stack simple is used. And it is cut from the top surface of sample using the ThermoFisher scios 2 FIB-SEM system, and then Pt film is deposited on the surface to protect perovskite film. Then, the sample is welded to the 3 mm diameter copper FIB bracket. Contacting the left side of sample with a tungsten tip; and the movement of tungsten tip is controlled by piezoelectric ceramics at a rate of about 0.01 nm/s to apply bending force on the edge of sample. In all bending procedures, a 300 kV voltage and a weak electron beam are employed in the TEM system to reduce the potential influence of the beam on the bending deformation. The camera captures the real-time stress distribution at a frame rate of 20 frames per second.

**Device characterizations**

*Photovoltaic performance characterization.* The J-V curves are obtained using a Keithley 2400 source measurement unit in conjunction with a solar simulator from RAVI SYSTEMS CORP. under AM1.5G illumination, utilizing a high spectral match solar simulator (RHS-50SS) equipped with a xenon lamp and a halogen lamp. The light in the short wavelength range was calibrated by 91150V-KG5 reference cell and then the entire light intensity was calibrated by 91150V reference cell. The tandem devices are exposed to light through a mask with an aperture area of 1.04 $cm^2$. The J-V curve measurements are conducted in an unencapsulated state in air. The external quantum efficiency (EQE) spectra are recorded with the Enli Technology QE-R system. -0.5 V bias is applied when measuring the current of perovskite subcells. For the maximum power point (MPP) tracking test, continuous illumination from a LED-based solar simulator, which provided a light intensity of 100 $mW/cm^2$ is conducted within a nitrogen atmosphere. Subsequently, PCE is measured normalized before aging.

*Bending Durability Test.* The flexible tandems are secured to two horizontal fixed poles and perform horizontal reciprocating motion by adjusting the distance between the poles, thereby testing various bending radius and cycles. The bending radius is calculated according to the method we describe in *Supplementary Fig.19.* And particularly to bending tests with different bending radii, we conduct 50 cycles of bending at each radius.

*Three-point Bending Test.* Place the flexible PSTs horizontally on two fixed poles, with one free pole exerting a vertical force. The force is gradually loaded to bend the samples. The force (F) and numerical displacement (D) applied by the free pole are record until the sample fractures.

**Density Functional Theory Calculation.** Spin-polarized density functional theory (DFT) calculations are performed using the Vienna Ab initio Simulation Package (VASP). The Perdew-Burke-Ernzerhof (PBE) generalized-gradient approximation is employed to characterize the electron-electron interactions. An energy cutoff of 400 eV is applied. The Monkhorst-Pack k-point grid is set to $1\times2\times2$ for all computational tasks. The convergence threshold for energy and force is set to be less than $10^{-5}$ eV and 0.02 eV/Å. Bottom layers are fixed during the calculations. All the structure is optimized.

**Acknowledgement**

This work is supported by National Key Research and Development Program of China (No. 2023YFB4202501), the National Natural Science Foundation of China (No. 62274026), the Sichuan Science and Technology Program (No. 2024NSFSC0216), the Fundamental Research Funds for the Central Universities (ZYGX2022YGRH010). W. Liu acknowledges the support from the National Natural Science Foundations of China (No. T2322028), Talent plan of Shanghai Branch, Chinese Academy of Sciences (No. CASSHBQNPD-2023–001), Shanghai Rising-Star Program (No. 23QA1411100).

## Author contributions

M.L., Z.L., F.L., Y.S., H.Z. and W.L. conceived and designed the overall project. Y.S. and H.Z. fabricated all the tandem devices and conducted the characterization. W.L. designed and fabricated the flexible silicon heterojunction solar cells with blunted edges. F.L. W.L. H.Z. Y.S. and J.D. designed and performed the bending test on the flexible tandems. L.Z., J.D., J.S., G.W., A.H., N.W. and F.M. helped with the fabrication of flexible silicon heterojunction solar cells and provided suggestion on bending tests of the flexible tandems. L.M., Q.L., Y.H. and T.Y. helped with the tandem device fabrication. L.M., Y.M. and Y.Q. helped with the material characterization and data analysis. X.S. performed the steady-state PL and CLSM measurements. M.L., Y.S., F.L. and H.Z. discussed the results and co-wrote the overall manuscript. Z.W. helped with writing and revision of the manuscript. M.L. and Z.L. supervised the project. All authors read and commented on the manuscript.

## Conflicts of Interest

The authors declare no competing interests.

## Data availability

The data that support the findings of this study are available from the corresponding author upon request.

## Supporting Information

# Flexible Perovskite/Silicon Monolithic Tandem Solar Cells Approaching 30% Efficiency

Yinqing Sun[1]†, Faming Li[1]†, Hao Zhang[1]†, Wenzhu Liu[2]†, Zenghui Wang[3], Lin Mao[1], Qian Li[1], Youlin He[1], Tian Yang[1], Xianggang Sun[1], Yicheng Qian[1], Yinyi Ma[1], Liping Zhang[2], Junlin Du[2], Jianhua Shi[2], Guangyuan Wang[2], Anjun Han[2], Na Wang[2], Fanying Meng[2], Zhengxin Liu[2]*, Mingzhen Liu[1,4]*

[1]School of Materials and Energy, University of Electronic Science and Technology of China, Chengdu 611731, P.R. China

[2]Research Center for New Energy Technology, Shanghai Institute of Microsystem and Information Technology, Chinese Academy of Sciences, Shanghai, P.R. China

[3]Institute of Fundamental and Frontier Sciences, University of Electronic Science and Technology of China, Chengdu 610054, P.R. China

[4]State Key Laboratory Electronic Thin Film and Integrated Devices, University of Electronic Science and Technology of China, Chengdu 611731, P.R. China

†These authors contributed equally to this work.

*Corresponding authors: z.x.liu@mail.sim.ac.cn; mingzhen.liu@uestc.edu.cn

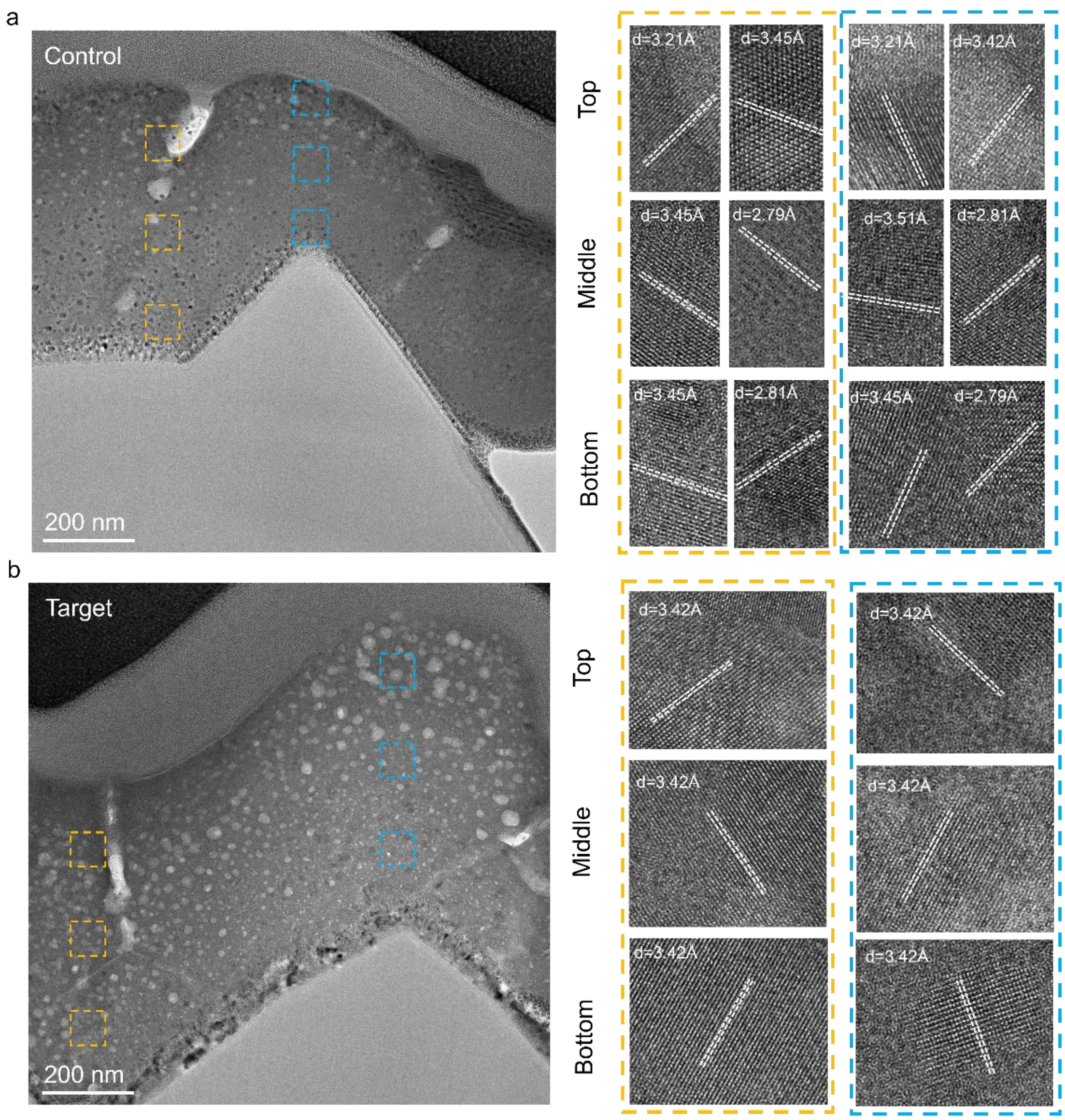

**Supplementary Fig. 1** | High-resolution TEM images collected from the corresponding squares of (**a**) control and (**b**) target film respectively.

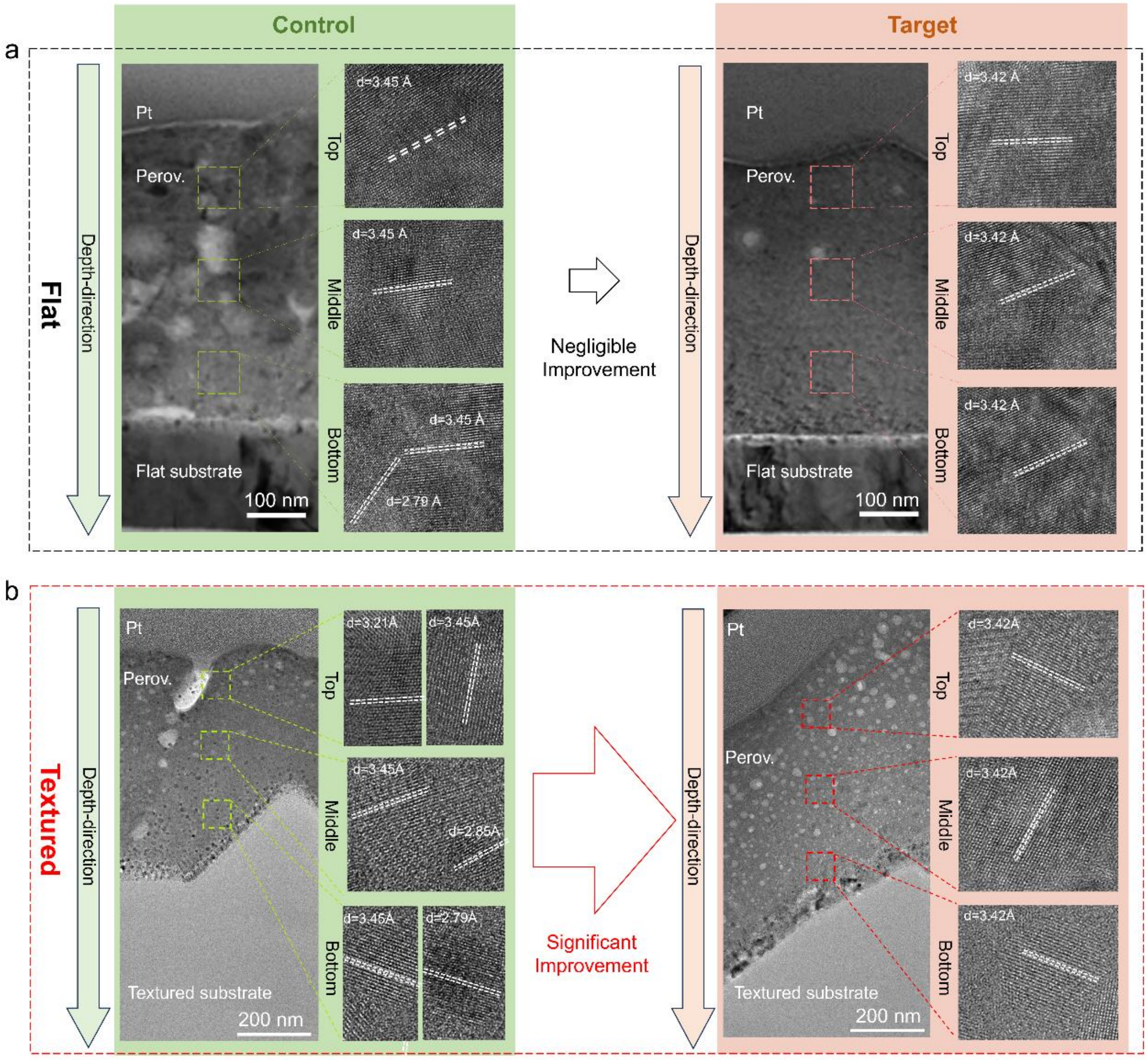

**Supplementary Fig. 2 |** High-resolution TEM images collected from the corresponding squares of control and target perovskite (Perov.) films on **a,** flat substrate and **b,** textured substrate, respectively.

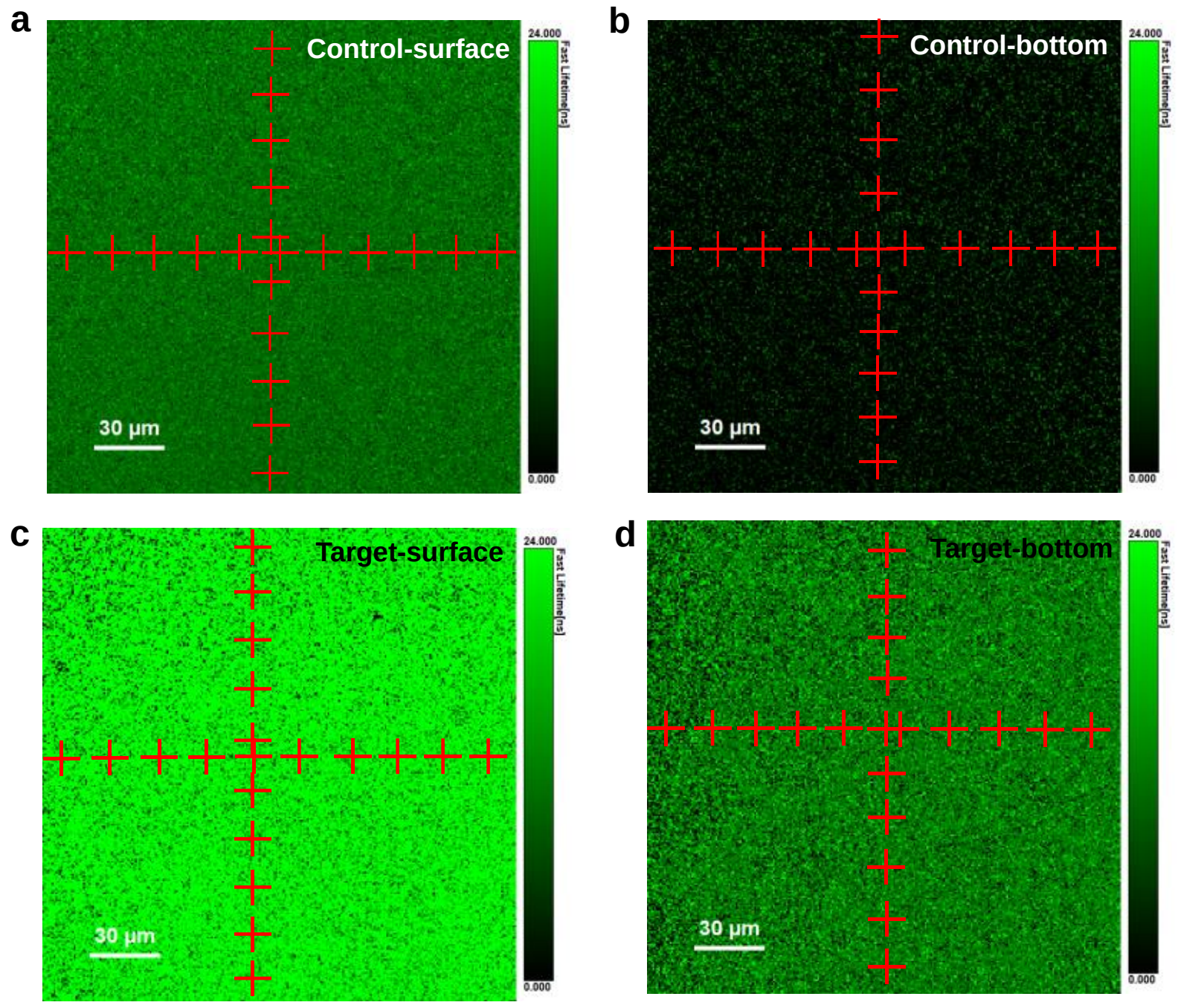

**Supplementary Fig. 3** | 20 points selected in CLSM from top and bottom surface of control and target film respectively for steady PL measurements.

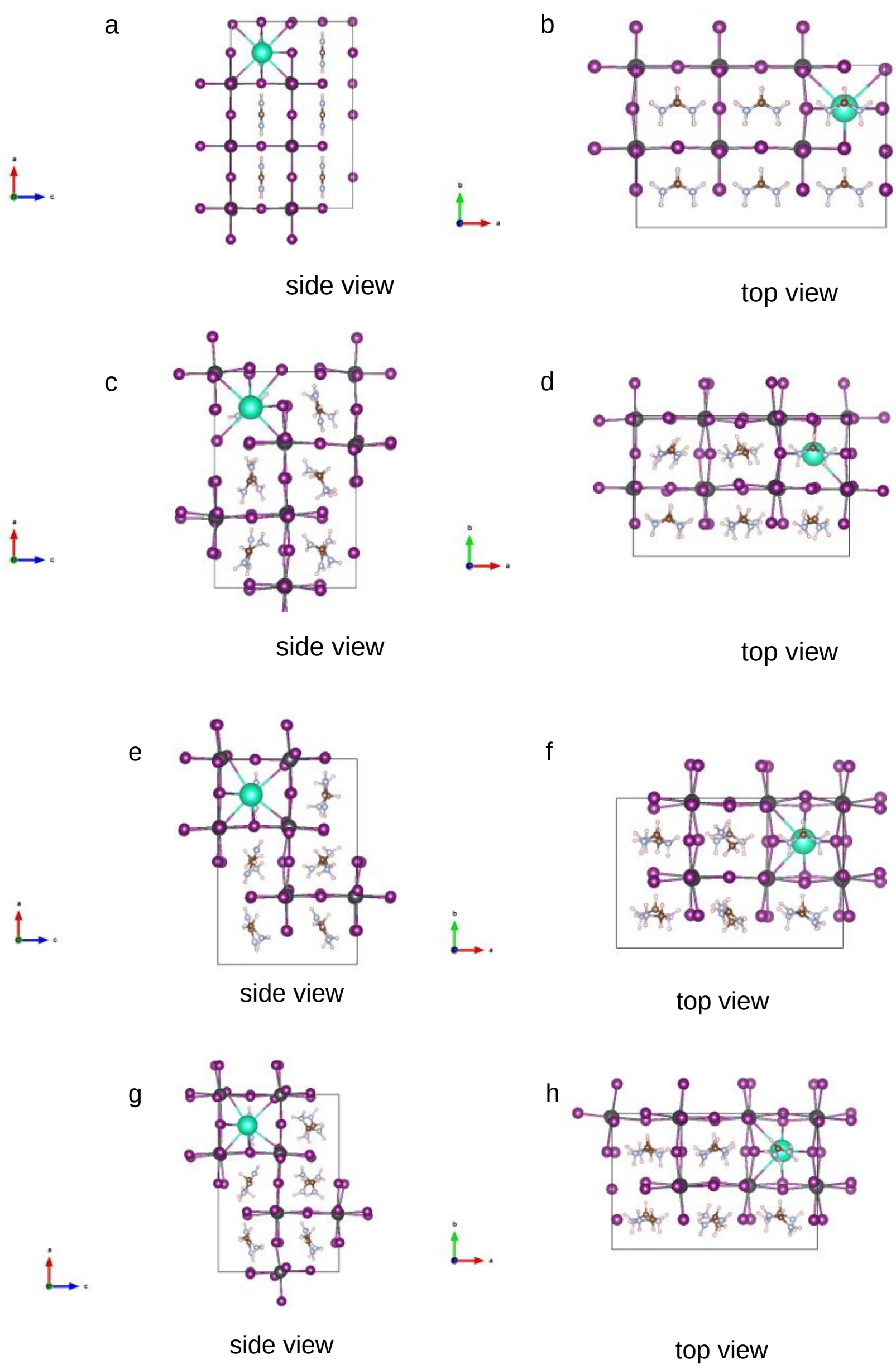

**Supplementary Fig. 4 | a, b,** Optimized structure for the structural energies of the single organic cation perovskite; **c, d,** Optimized structure for structural energies of the perovskite with a single MA cation substitution. The difference of these two structural energies is indicative of the formation energy of the control perovskite. **e, f,** Optimized structure for the structural energies of the mixed organic cation perovskite; **g, h,** Optimized structure for structural energies of the perovskite with a single FA cation substitution. The difference of these two structural energies is indicative of the formation energy of the target perovskite.

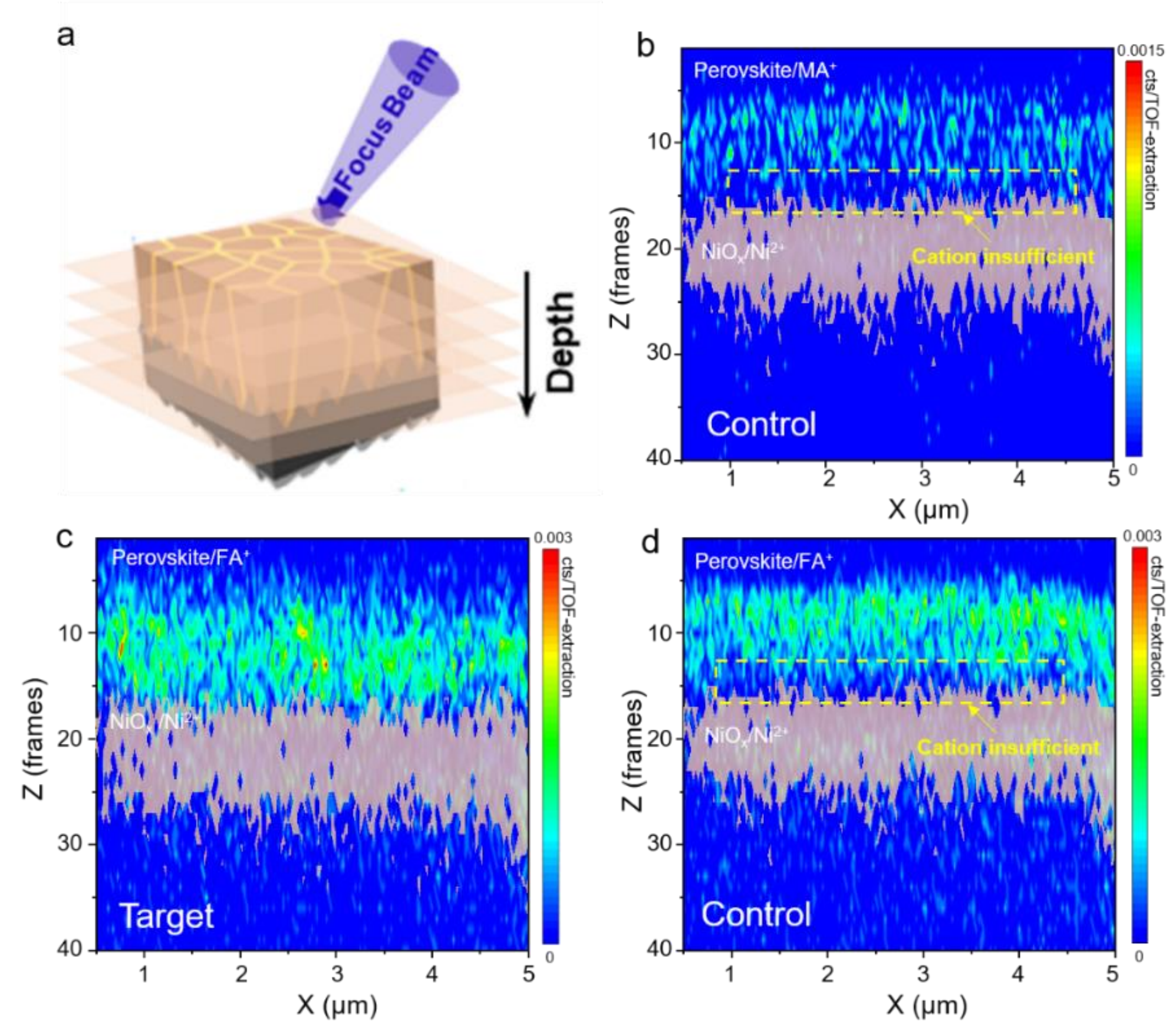

**Supplementary Fig. 5 | a,** Schematic diagram of FIB-ToF-SIMS test; **b-d,** Sideview of FIB-ToF-SIMS results. (**b**) $MA^+/Ni^{2+}$ distribution in control film; (**c,d**) $FA^+/Ni^{2+}$ distribution in target and control film, respectively.

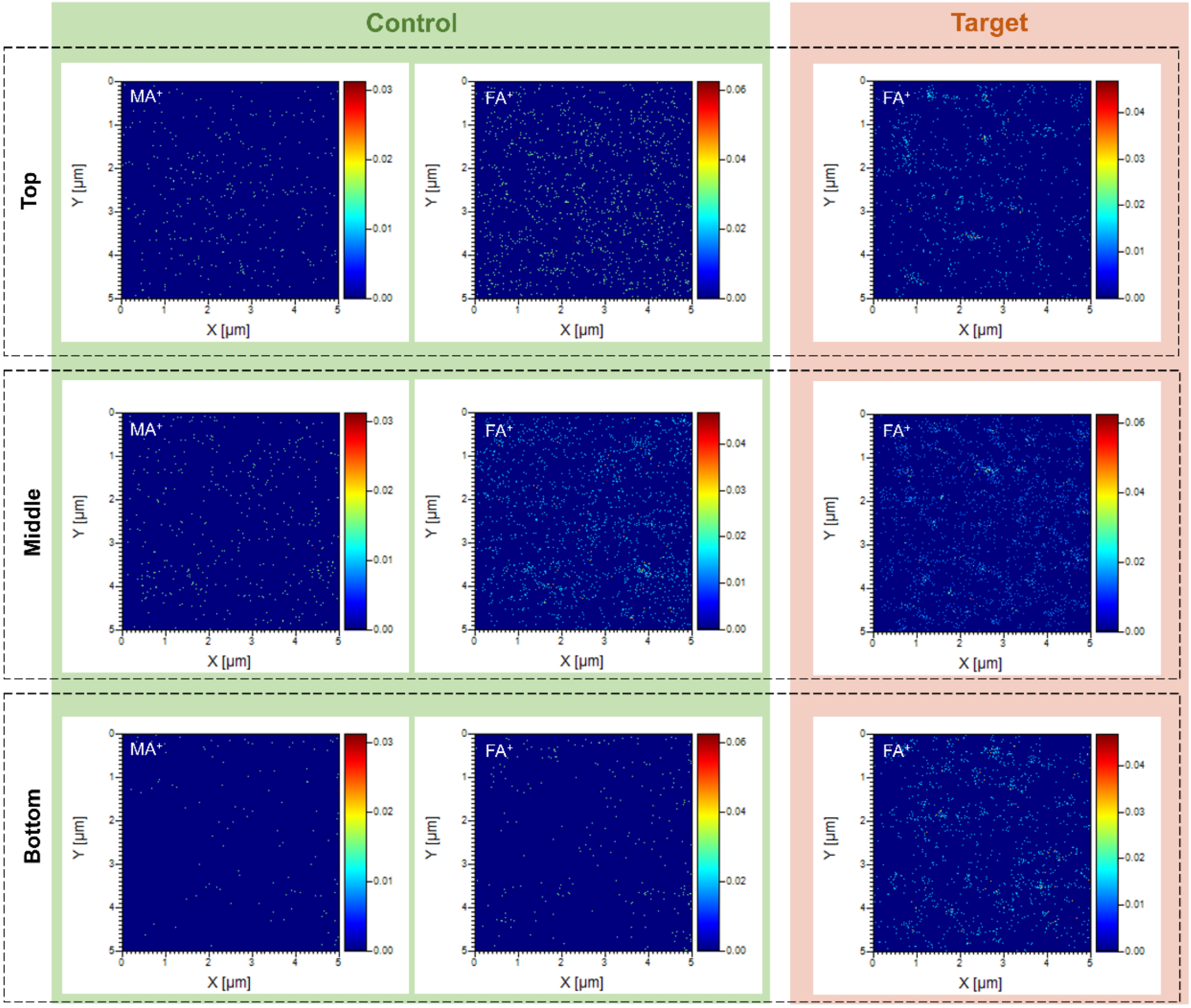

**Supplementary Fig. 6** | Top-view images of $MA^+$ and $FA^+$ distribution in Top, Middle and Bottom regions within control film and target film, respectively, sliced from the FIB-ToF-SIMS results.

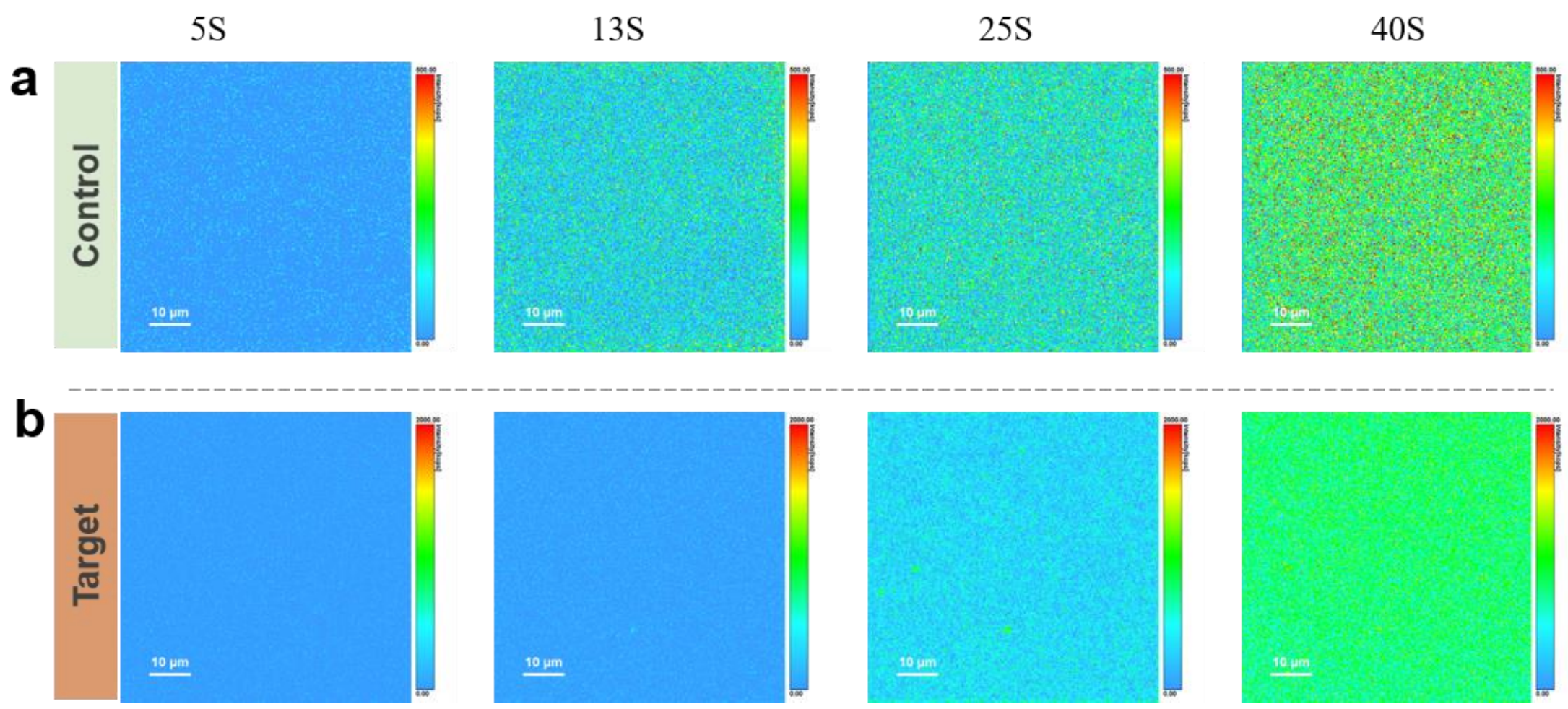

**Supplementary Fig. 7 | Real-time in-situ confocal laser scanning microscopy images**. **a,** control and **b,** target perovskite film on planer substrates.

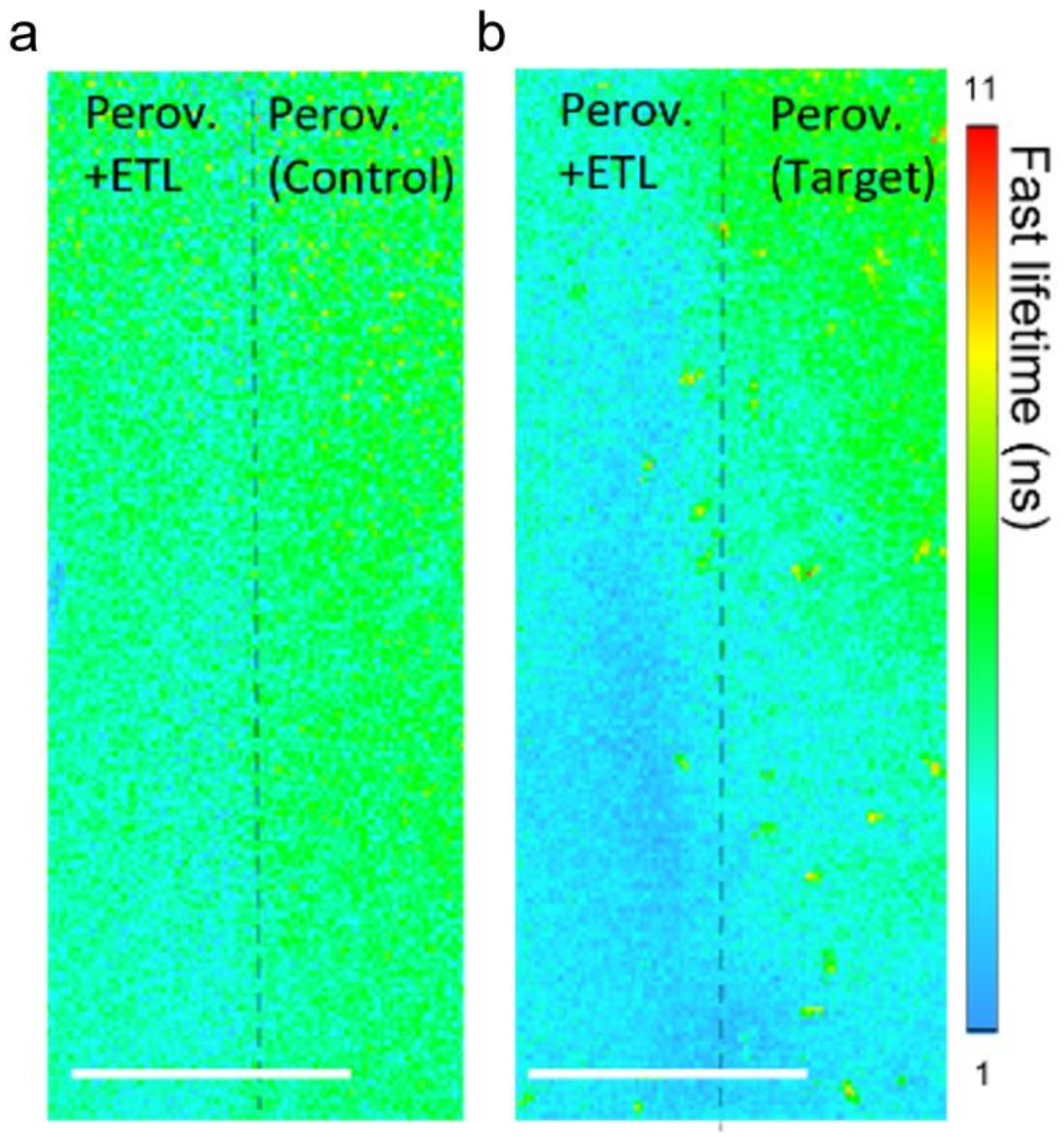

**Supplementary Fig. 8** | Direct observation of photoluminescence quenching of **a,** control and **b,** target perovskite (Perov.) films by the extraction effect of ETL (scale bars: 100 μm).

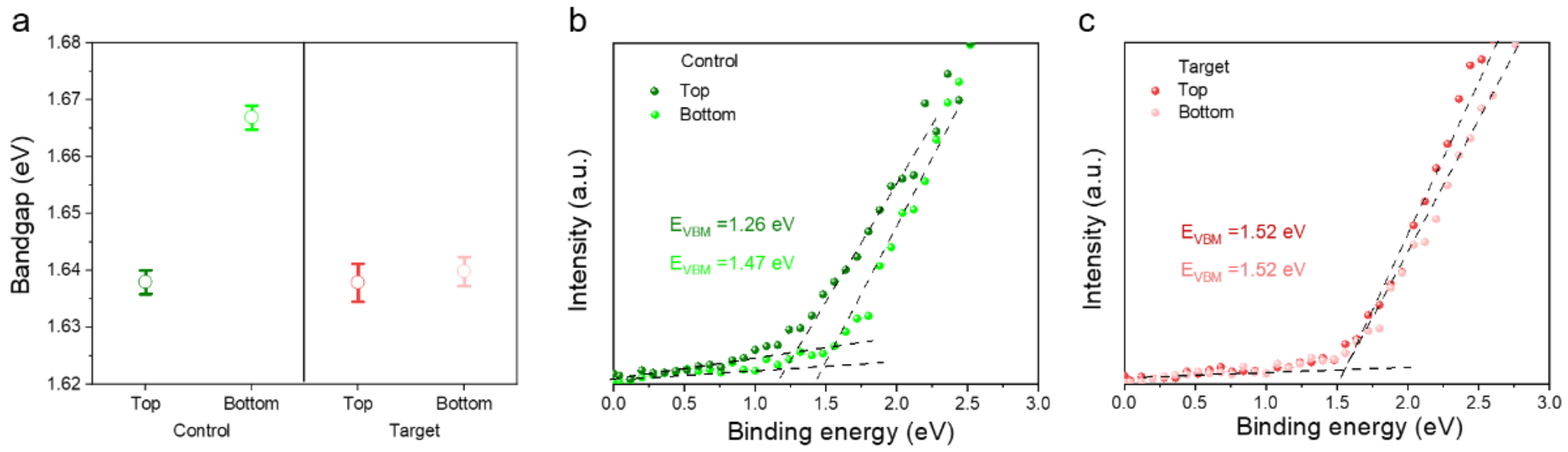

**Supplementary Fig. 9 | a,** Interval plots of the wavelength of PL emission peak (all the 20 spectra are averaged) of Top and Bottom region of control and target perovskite films, draw from Fig. 2c. UPS spectra displaying the valance band of Top and Bottom region of **b,** control and **c,** target perovskite films.

Regarding "the process for calculating the energy level", we detailed below:

1. Determine the energy level of $E_F$:

   We first determine the Fermi level with respect to the vacuum level. Calculating the work function ($W_F$) using following formula:

$$W_F = hv - E_{cutoff}$$

   where $hv$ (21.22 eV) is the energy of incident photon, $E_{cutoff}$ is the cut-off energy of photoelectron spectra, and the $E_{cutoff}$ before equilibrium of Control-Top, Control-Bottom, Target-Top, and Target-Bottom are 17.04, 16.56, 17.10, and 17.05 eV, respectively. $E_F$ is the energy of the Fermi level, which equals to $E_{vac}$ minus $W_F$.

2. Determine the energy of the valence band maximum ($E_{VBM}$):

   We next determine the energy level of the valence band, by identifying the starting point of the photoelectron spectra (Supplementary Figs. 9b and 9c), which corresponds to the energy difference from the $E_F$ to $E_{VBM}$.

3. Determine the bandgap ($E_g$) and energy of conduction band minimum ($E_{CBM}$):

   We next determine the energy level of the valence band, by calculating $E_g$ of perovskites from PL results (Supplementary Fig. 9a) using the following formula:

$$E_g = \frac{hc}{\lambda_{peak}}$$

where $h$ is the Planck constant, $c$ is the speed of light and $\lambda_{peak}$ is he wavelength of corresponding PL emission peak. Based on that, we obtain the $E_{CBM}$, by adding $E_{VBM}$ and $E_g$.

4. Determine the energy level structure of our perovskites:

The energy level structure of our perovskite films is obtained through the combination of the calculated $E_F$, $E_{VBM}$ and $E_{CBM}$[1-3].

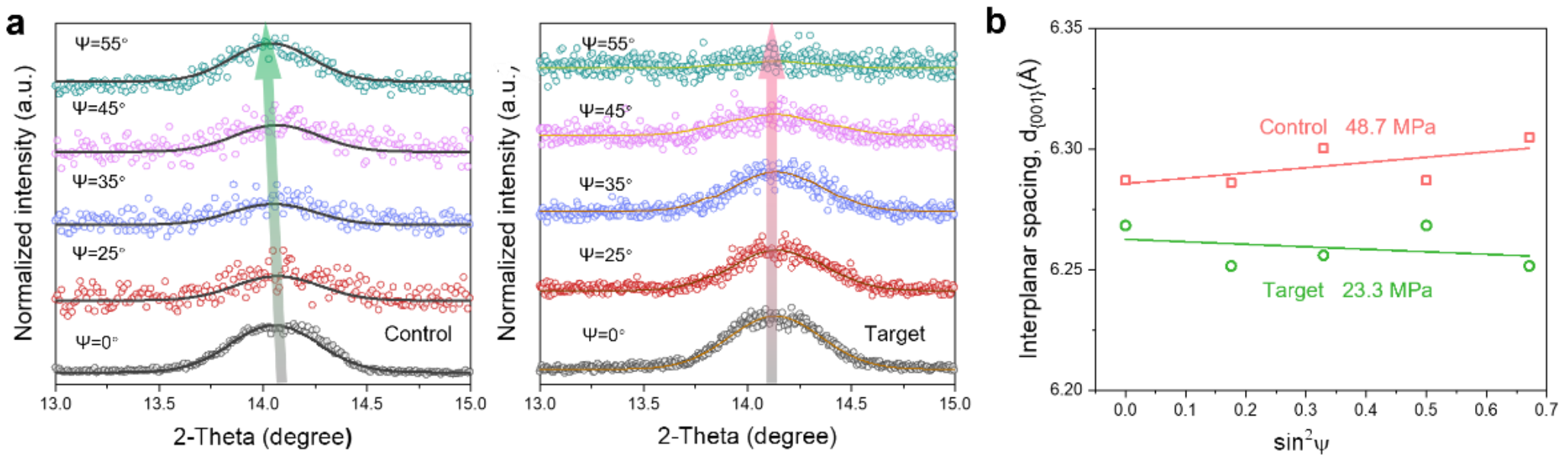

**Supplementary Fig. 10** | GIXRD patterns at different Ψ angles (from 0 to 55°) of **a,** control and **b,** target perovskite films respectively. **c**, Linear fitting of lattice spacing $d_{(001)}$-$\sin^2\varphi$ for two types perovskite films.

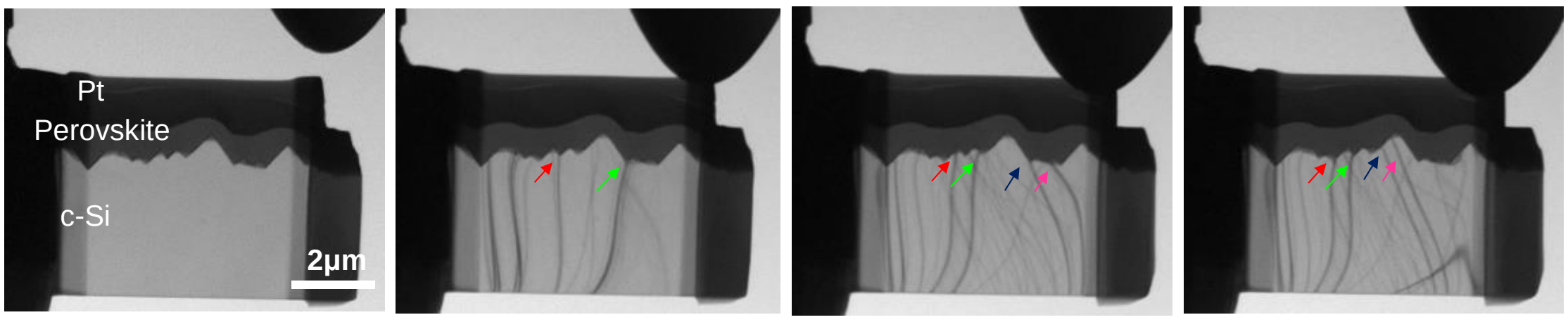

**Supplementary Fig. 11** | TEM images of in-situ bending test performed on c-Si/ITO/HTL/perovskite half stack. (a tungsten tip controlled by the piezo manipulator is used to load force at the right side of the FIB perovskite/silicon sample)

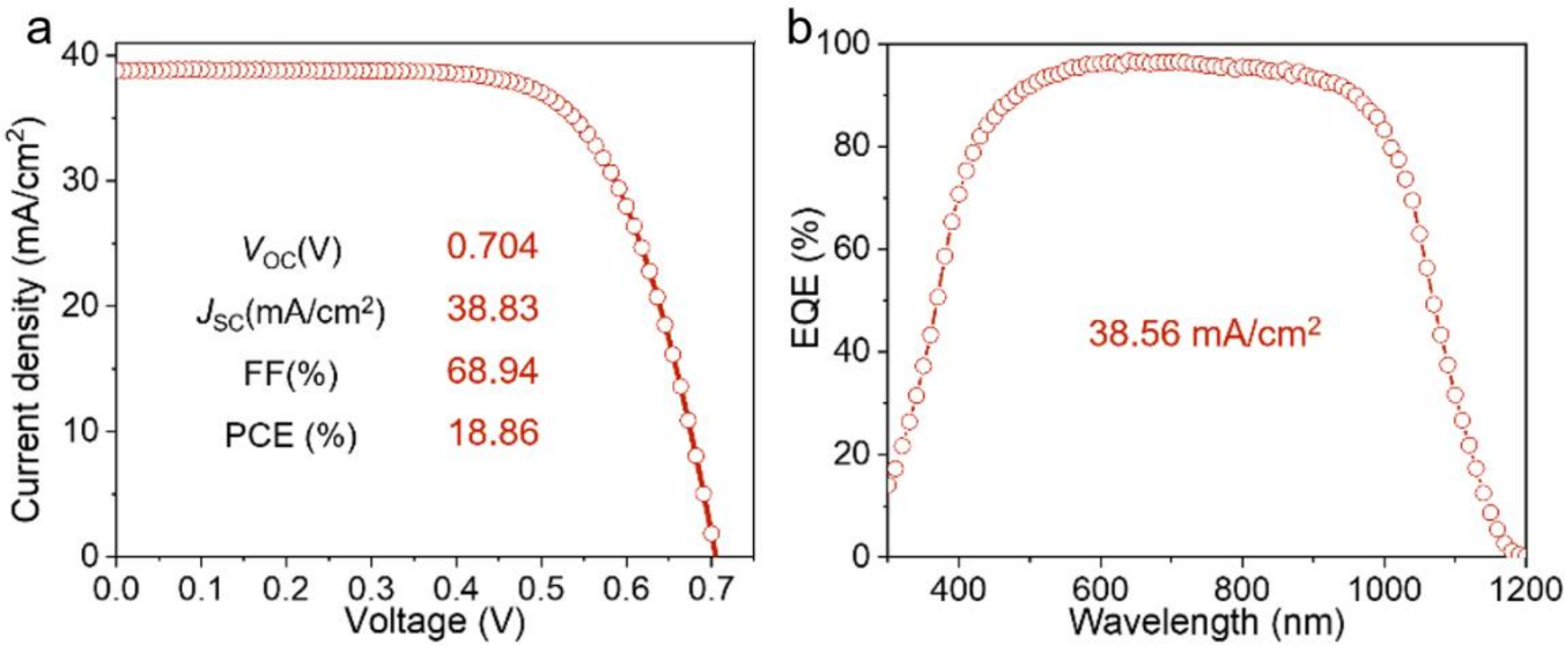

**Supplementary Fig. 12 | a**, J-V curves and **b,** EQE curves of 70 μm crystalline silicon single-junction solar cell.

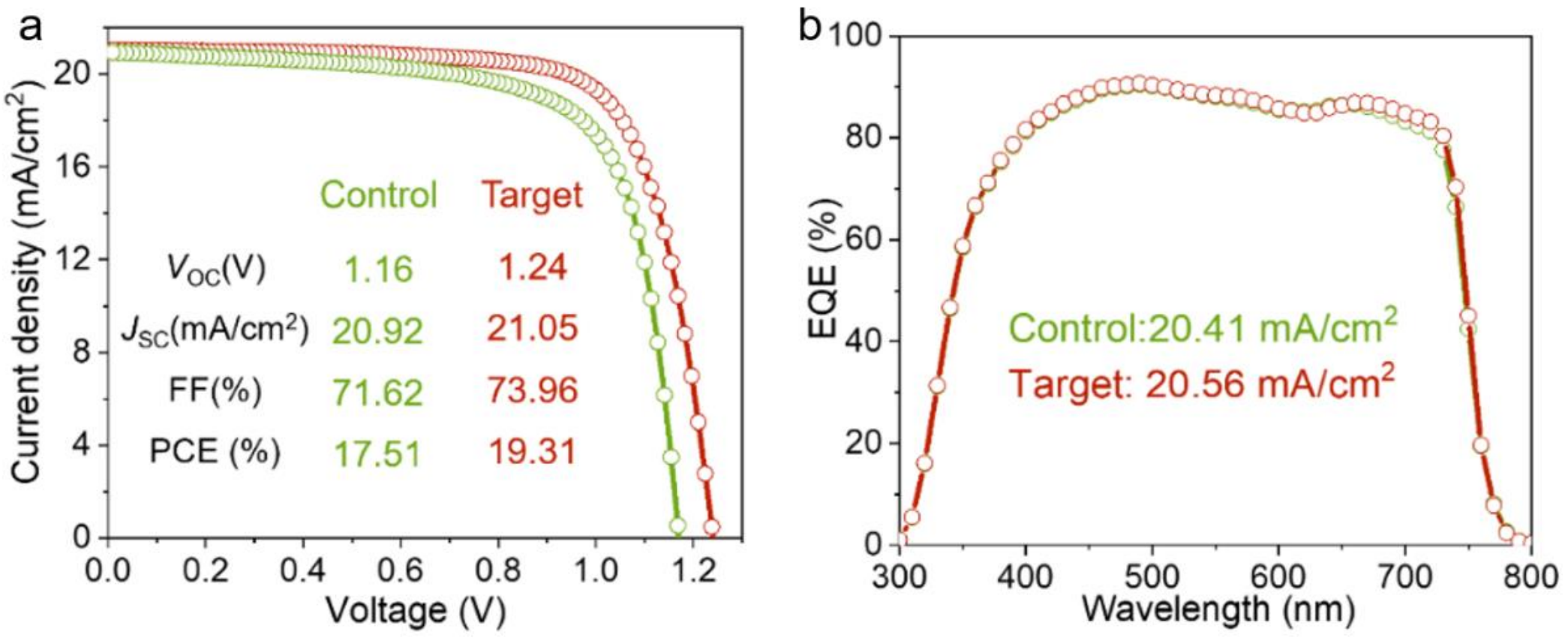

**Supplementary Fig. 13 | a,** J-V curves and **b,** EQE curves of control and target single-junction perovskite solar cells based on flat ITO glass.

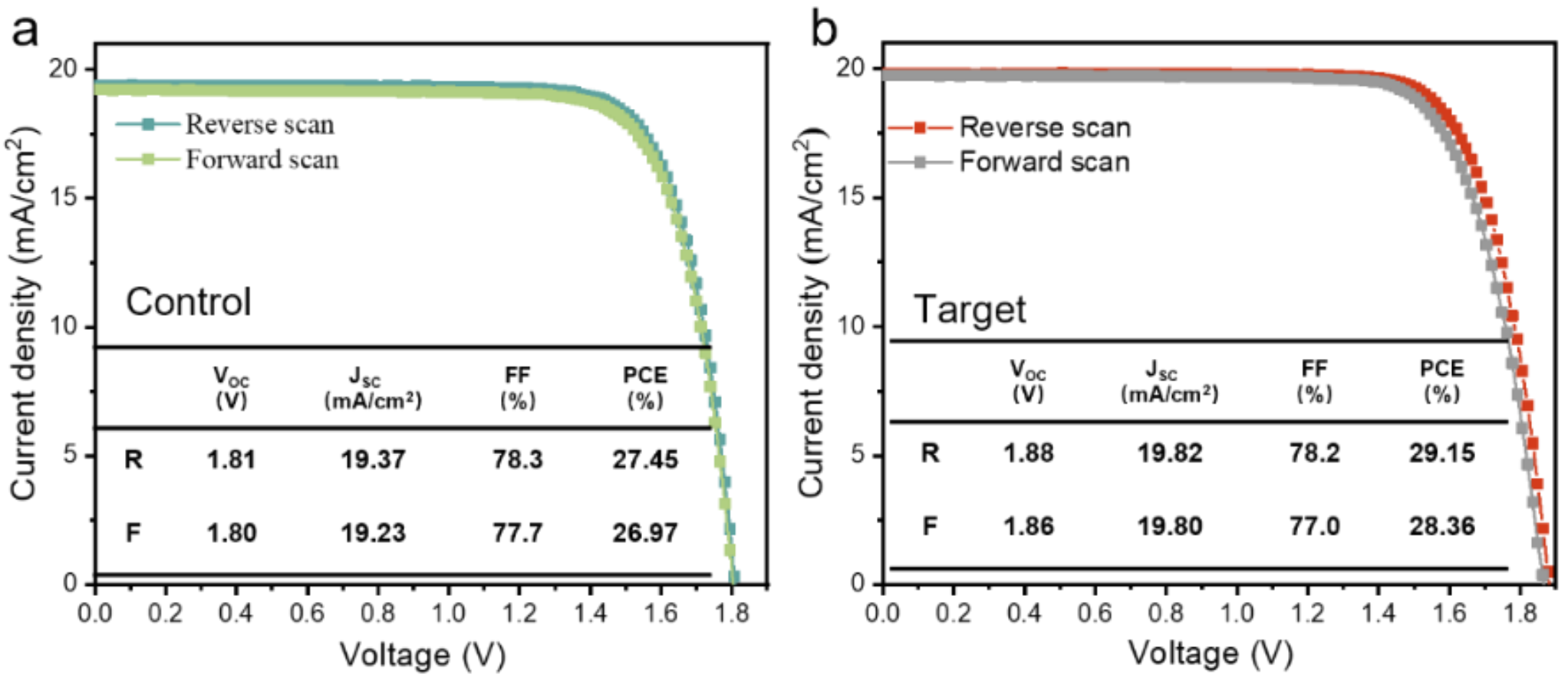

**Supplementary Fig. 14 |** The J-V curves and PV parameters (inset) of **a,** control and **b,** target flexible perovskite/c-Si tandem solar cells.

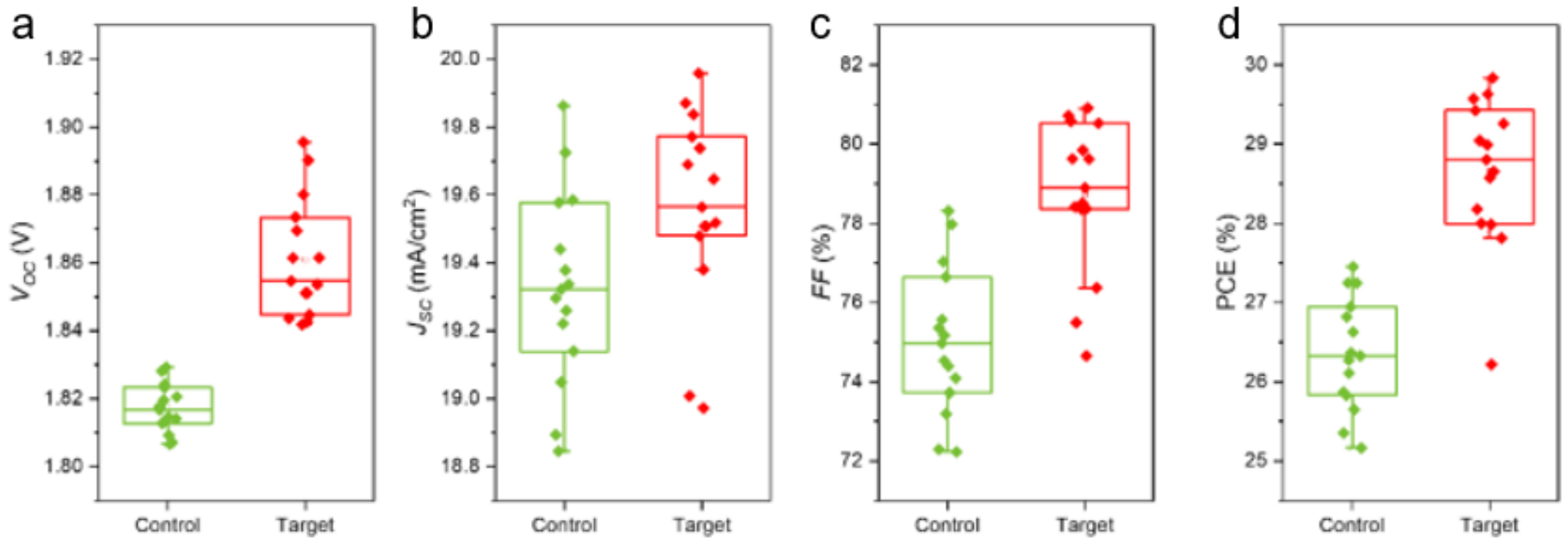

**Supplementary Fig. 15** | Box plots of photovoltaic parameters of control and target flexible tandem devices.

100027826

(10) 23-11-23

# 检 测 报 告

Test Report

ilac-MRA CNAS 中国认可 国际互认 检测 TESTING CNAS L0893

报告编号：检测字第 202311000550 号
Report No.

防伪码
ea15332ee8b82a6a
9f8794b733297be3
6ae8dde1a8f74822
7ddc90fede55019e

| | |
|---|---|
| 样品名称 Sample Name | 钙钛矿/薄硅柔性叠层太阳能电池 Perovskite/thin silicon flexible tandem solar celll |
| 标称生产单位 Manufacturer | 电子科技大学 University of Electronic Science and Technology |
| 委托单位 Client | 电子科技大学 University of Electronic Science and Technology |
| 联络信息 Contact Information | 成都市高新区（西区）西源大道 2006 号 No.2006, Xiyuan Ave, West Hi-Tech Zone, Chengdu |
| 检测类别 Test Category | 委托检测 commission detection |

中国测试技术研究院 检测专用章 (检测专用章) Stamp

扫码 验真

1003660035

授权签字人 Approved by 陈潇潇

签发日期 Issue Date 2023 年 Year 11 月 Month 23 日 Day

地址：中国·四川·成都玉双路 10 号
Address: No.10, Yushuang Road, Chengdu, Sichuan, China
邮编：610021
Post Code
网址：www.nimtt.cn
Web
电话：028-84404337
Telephone
传真：028-84404149
Fax
邮箱：kfzx@nimtt.com
E-mail

# 检 测 结 果

Results of Test

**1.检测条件（Test Condition）**

标准太阳电池：硅太阳电池(KG1 玻璃)

Reference Cell: Mono-Si Solar Cell (Window Material: KG1)。

被测样品信息：钙钛矿/薄硅柔性叠层太阳能电池：尺寸（2×2）$cm^2$，孔径面积:1.04$cm^2$

Sample Information:Perovskite/thin Silicon Flexible Tandem Solar Celll with Size (2×2) $cm^2$ and an Aperture Area of 1.04 $cm^2$。

样品在测试前的存放条件：温度 25±5℃，相对湿度 25±5%，避光保存 1 天。

Storage Condition of Sample Before Test: Temperature: 25±5℃; Humidity: 25±5%; stored in Dark for 1 Day。

**2.检测方法及参数设置(Methodologies and Settings)**

(1) 在 3A 级稳态太阳模拟器下（AM1.5 光谱），首先用标准太阳电池标定太阳模拟器辐照度至 1000 W/$m^2$，通过水浴恒温台控制电池片温度 25℃，再采用数字源表对被测样品进行 *I-V* 参数测量。

In the 3A steady-state solar simulator (Spetrum:AM1.5), the irradiance of the solar simulator was first calibrated to 1000 W/$m^2$ with a standard solar cell, the temperature of the battery was controlled by a water bath thermostat at 25℃, and then the I-V parameters of the tested sample were measured with a digital source meter.

(2) 测量软件参数设置如表 1 所示：

The parameter Settings of the measurement software are shown in Table 1:

表 1 测量参数设置(Table 1 Parameter Settings for I-V Test)

| 扫描类型 Scan Mode | 起始电压(V) Start Voltage | 终止电压(V) End Voltage | 扫描间隔(V) Step | 扫描点延时(ms) Sweep point delay | 扫描点数 Point | 预光照 Light Soaking Pre-treatment |
|---|---|---|---|---|---|---|
| 正扫 Forward scan | -0.1 | 1.95 | / | 10 | 150 | NO |
| 反扫 Reverse scan | 1.95 | -0.1 | / | 10 | 150 | NO |

**3.检测结果 (Test Results)**

电流-电压特性曲线如图 1、图 2 所示：

Current-Voltage Curves are Shown in Figure 1 and 2.

| 备注 Note | / |
|---|---|

# 检 测 结 果

Results of Test

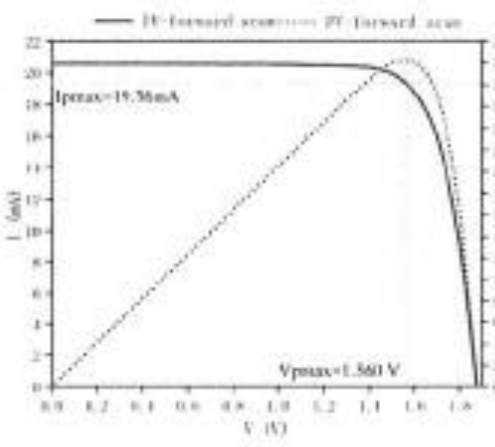

图 1 正扫 I-V 曲线

Figure 1 I-V Curve (Forward Scan)

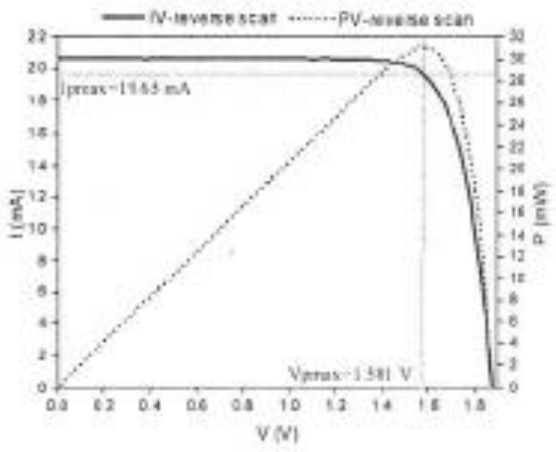

图 2 反扫 I-V 曲线

Figure 2 I-V Curve (Reverse Scan)

表 2 I-V 特性参数

Table 2 Irradiated I-V parameters

| 扫描类型 Scan Mode | 短路电流 Short-circuit Current | 开路电压 Open-circuit Voltage | 填充因子 Fill Factor | 最大功率 Maximum-Power | 最大功率点电压 Maximum-Power Voltage | 最大功率点电流 Maximum-Power Current | 转换效率 Conversion Efficiency |
|---|---|---|---|---|---|---|---|
| | $I_{sc}$ (mA) | $V_{oc}$ (V) | FF (%) | $P_{max}$ (mW) | $V_{pmax}$ (V) | $I_{pmax}$ (mA) | η (%) |
| 正扫 Forward scan | 20.56 | 1.873 | 78.41 | 30.205 | 1.560 | 19.36 | 29.04 |
| 反扫 Reverse scan | 20.65 | 1.882 | 79.96 | 31.071 | 1.581 | 19.65 | 29.88 |

备注 Note
1. Reported Performance Parameters Take the Average of Three Test Values.
2. The battery area data is determined according to the mask plate area.

审核人员 Verified by 吴伟钢　　主检人员 Tested by 康 张 李

| No. | Z2412WT8888-022439-Y |
|---|---|
| Total page | 10 |

# TEST REPORT

Partial copying without authorization is prohibited

| | |
|---|---|
| Product Name : | Flexible perovskite/c-silicon monolithic tandem solar cells |
| Type and Specification : | / |
| Test Category : | Entrusted Test |
| Factory : | University of Electronic Science and Technology of China |
| Client: | University of Electronic Science and Technology of China |

中国赛宝实验室 检验检测专用章 (01)

China Electronic Product Reliability And Environmental Testing Research Institute

CEPREI

China CEPREI Laboratory

29462

## Detection Result

| No. | Test Items | Unit | Test conditions or Requirement | Test Result |
|---|---|---|---|---|
| 1. | I-V characteristic | % | Fix the sample onto a solar simulator with a light intensity of 1000 W/m² and a scanning range from -0.1 V to 1.95 V. Perform scans in two directions (from Isc to Voc and from Voc to Isc) during a single illumination period based on IEC 60904-1:2006 | Efficiency (Jsc to Voc): 28.94%; Efficiency (Voc to Jsc): 29.63% |
| 2. | Steady state output power | mW | Fixing the solar cell at the voltage of the maximum power point (determined by I-V characteristics) and tracking the change in current output for 300 seconds, the Pmax in the report is marked as the average value of the 300 seconds | 30.36mW |
| 3. | Thickness | μm | Using digital micrometer to measure the thickness of the device | 70μm |
| 4. | Sample bending characteristics | / | Applying external force at device for bending | Flexible |
| 5. | Aperture area | $mm^2$ | Using measuring microscope to define the length and width of aperture area of the mask and multiplying them together to get the area | 104.0 $mm^2$ |

## Table 1. Test Data

| Test Items | Unit | Forward Scan (Isc to Voc) | Reverse Scan (Voc to Isc) |
|---|---|---|---|
| Mask Area | $mm^2$ | 104.0 | |
| Voc | V | 1.938 | 1.915 |
| Isc | mA | 20.52 | 20.62 |
| Jsc | $mA/cm^2$ | 19.74 | 19.83 |
| Pmax | mW | 30.10 | 30.82 |
| Vmax | V | 1.590 | 1.603 |
| Imax | mA | 18.93 | 19.22 |
| Fill Factor | % | 75.65 | 78.05 |
| Efficiency | % | 28.94 | 29.63 |

TEST PHOTO

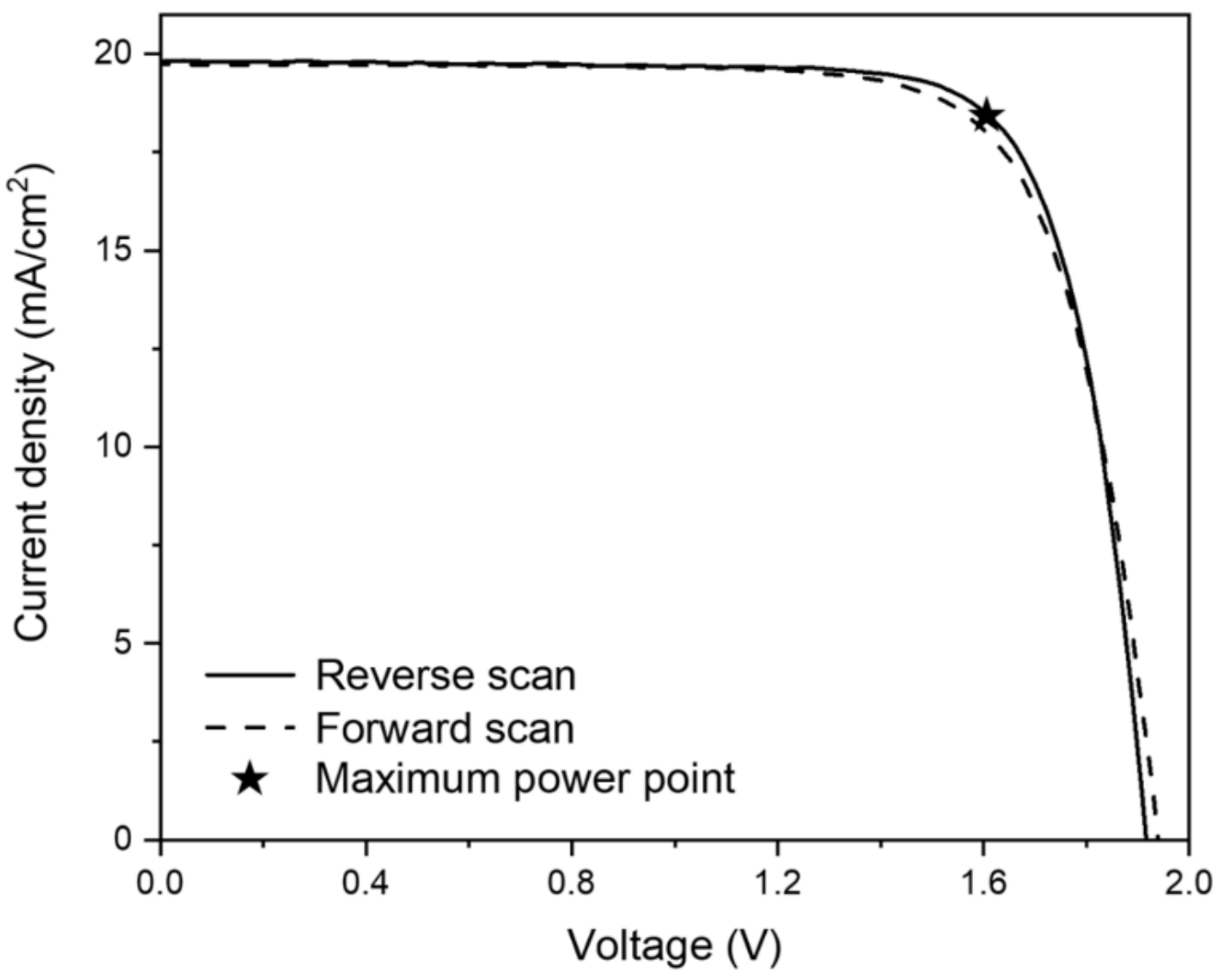

Figure 1. J-V curve

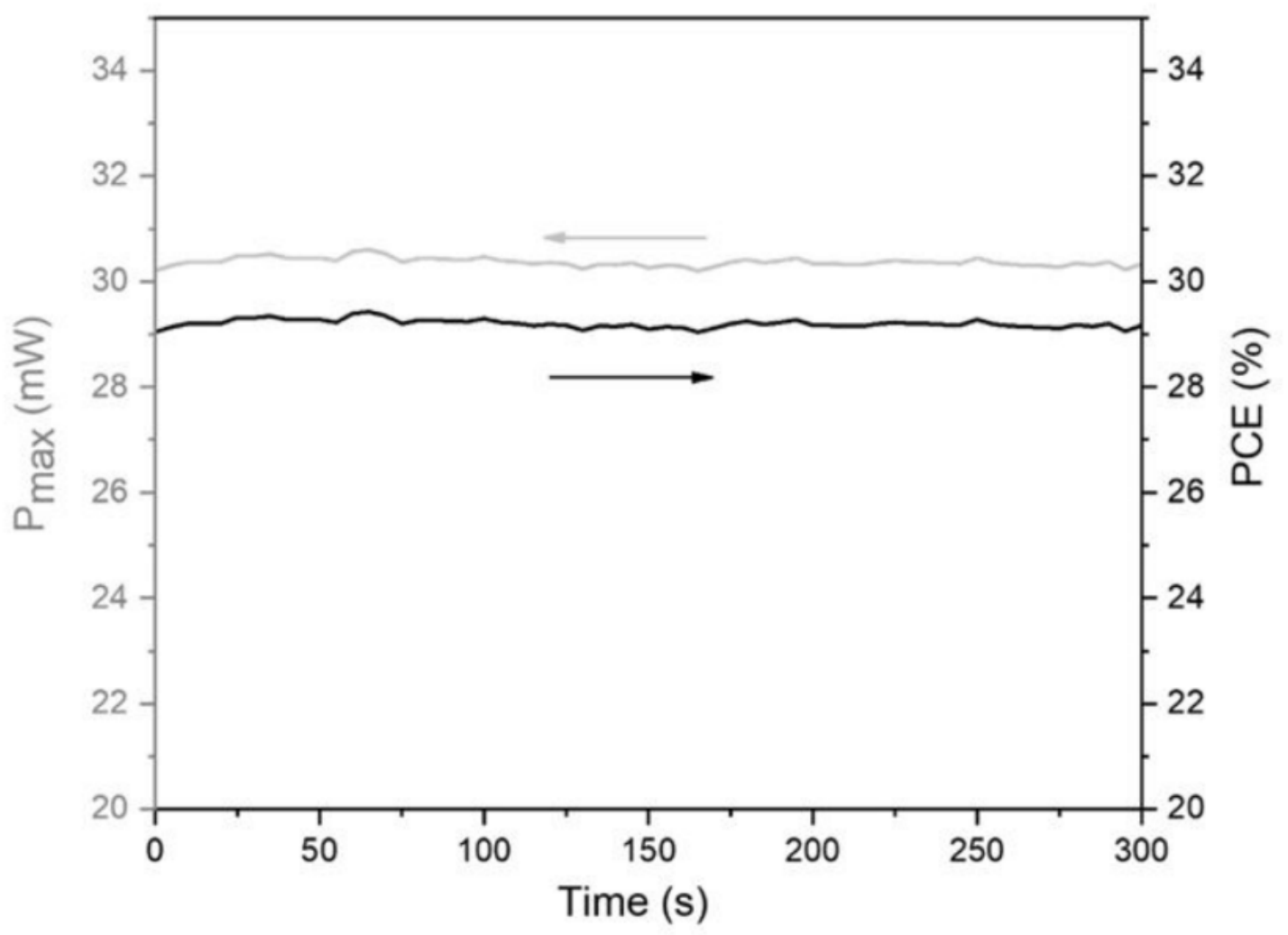

Figure 2. Steady-state scanning curve. ($V_{bias}$=1.603V, $t_{scanning}$=300s, stabilized $P_{max}$=30.36mW, stabilized *PCE*=29.20%)

TEST PHOTO

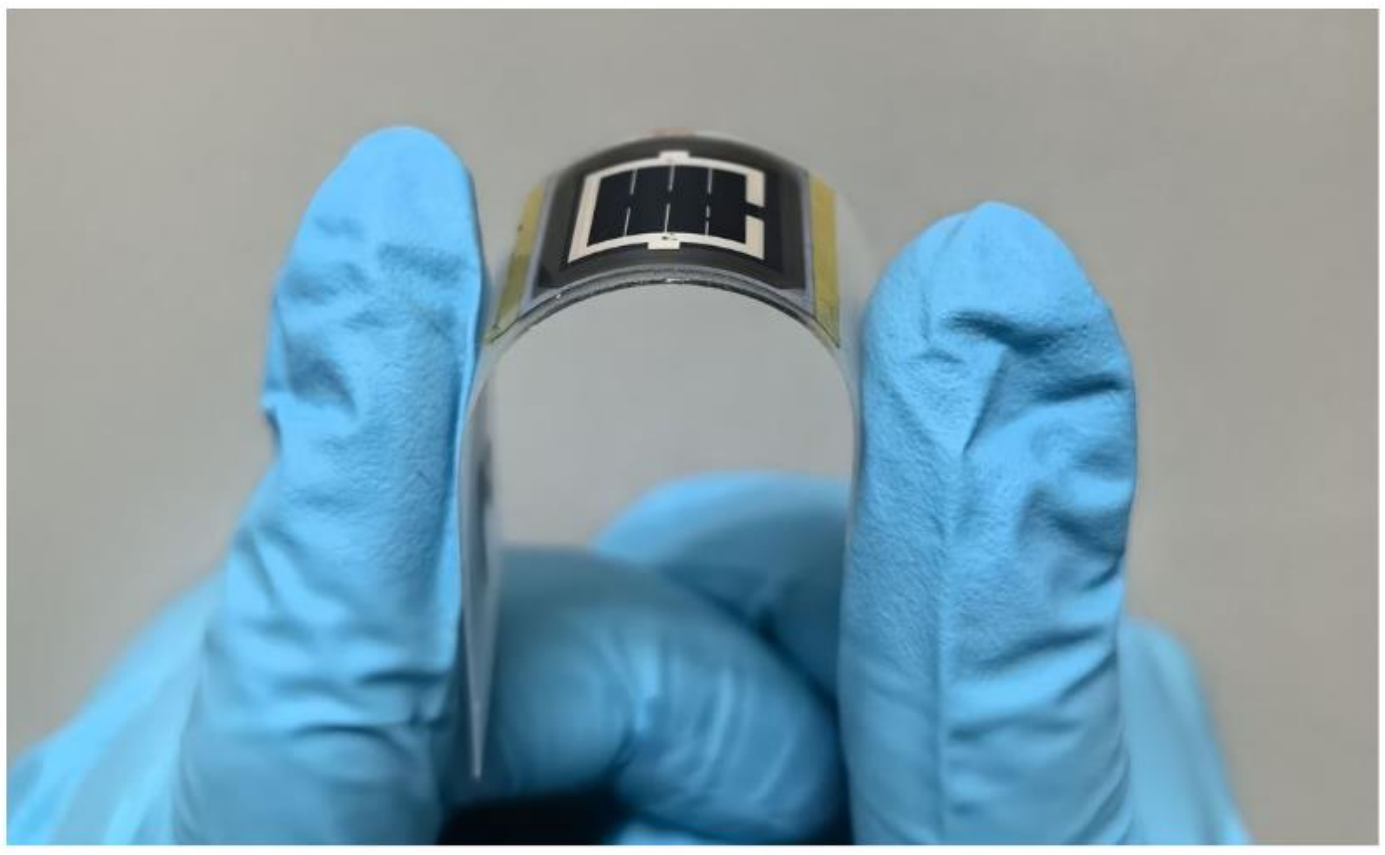

Figure 3. The sample can be bent

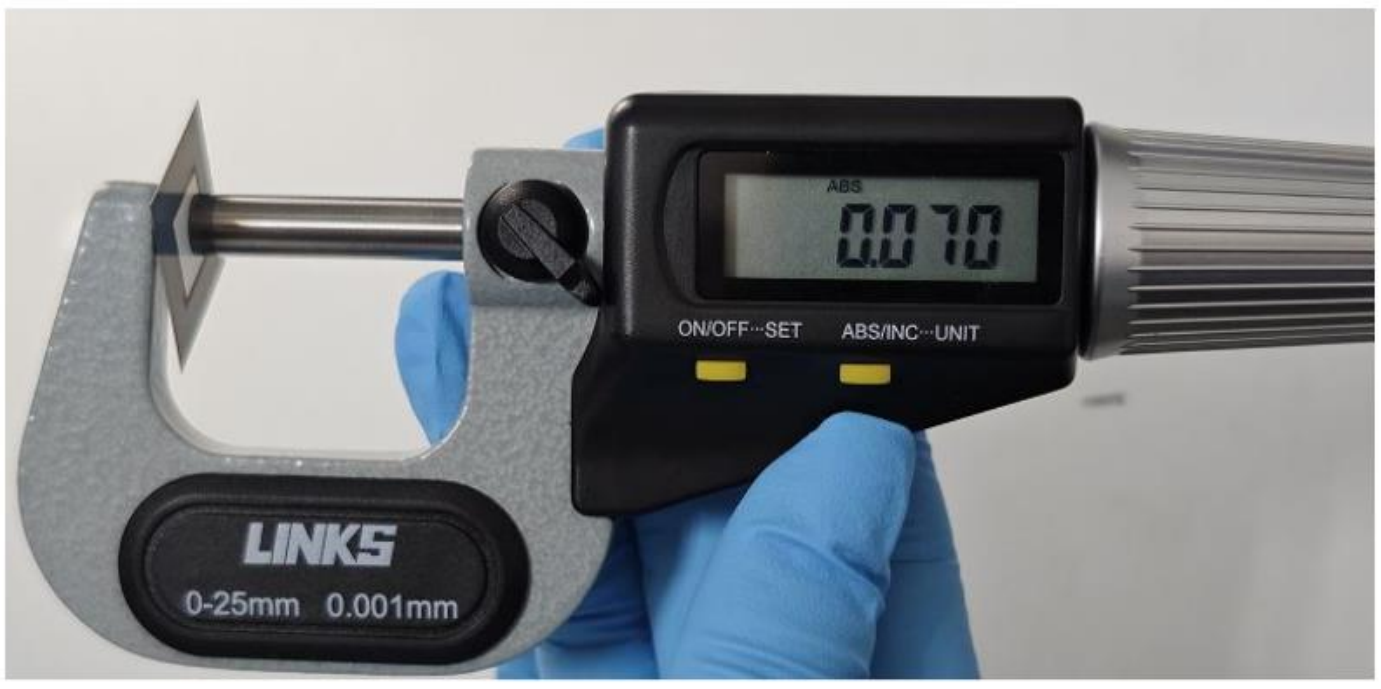

Figure 4. Sample thickness: 70±1μm

## List of test equipments

| No. | Test equipment | Model | Certificate No. | Date of Calibration |
|---|---|---|---|---|
| 1. | A+ A+ A+ Dual light source Solar Simulator | RHS-50SS | 202412108411 | 2024.12.30-2025.12.29 |
| 2. | Digital Source Meter | Keithely 2400 | HA5A2GD12301352 | 2024.12.30-2025.12.29 |
| 3. | Reference Cell | Mono-Si, WPVS (Callibration Value: 130.1mA) | 24Q2-01406 | 2024.08.16-2025.08.15 |
| 4. | Digital Micrometer | LINKS | HA5A2GD12310080 | 2024.12.31-2025.12.30 |
| 5. | Measuring Microscope | Leica S9i | HA5A2GD12310053 | 2024.12.31-2025.12.31 |

**Supplementary Fig. 16 |** Independent certifications from National Institute of Measurement and Testing Technology (NIMTT, Chengdu, China) and China Electronic Product Reliability and Environmental Testing Research Institute (CEPREI, Guangzhou, China) for our flexible perovskite/c-Si monolithic tandem solar cell, confirming a power conversion efficiency of 29.88% and 29.63% under AM1.5G illumination, respectively.

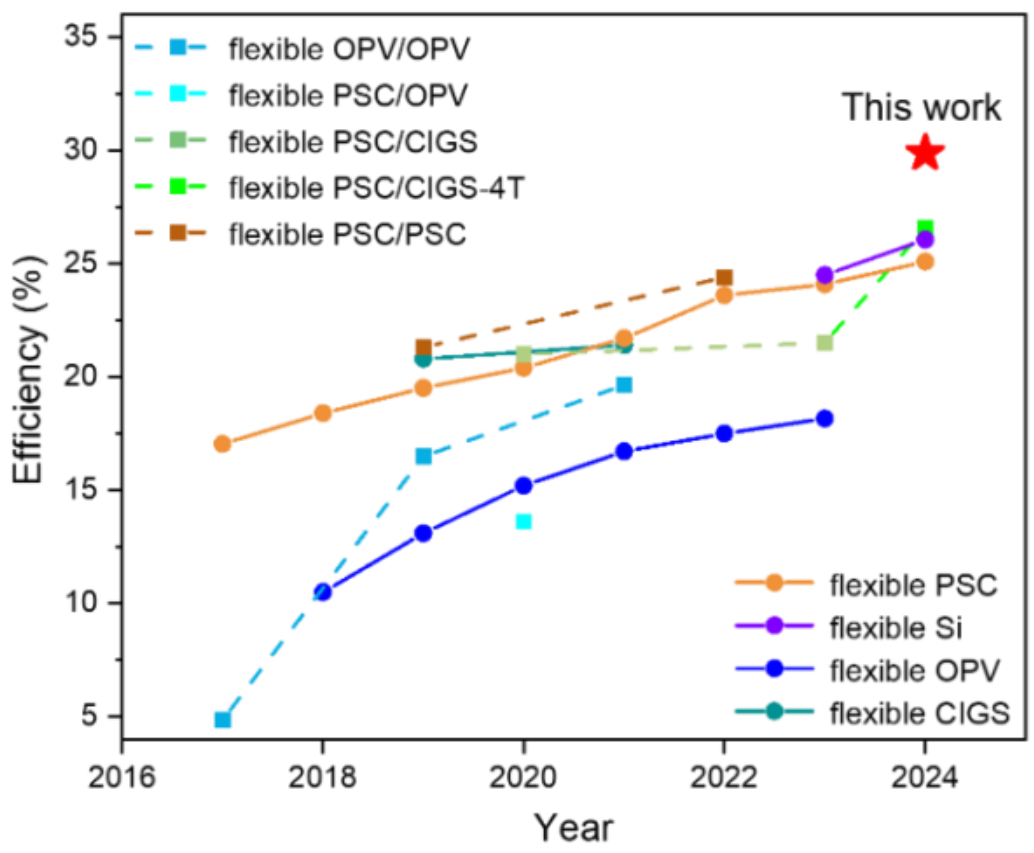

*CIGS: copper indium selenide; OSC: organic solar cell; PSC: perovskite solar cell; Si: silicon heterojunction solar cell*

**Supplementary Fig. 17 |** Summary of the efficiencies of flexible solar cells in recent years[4-24].

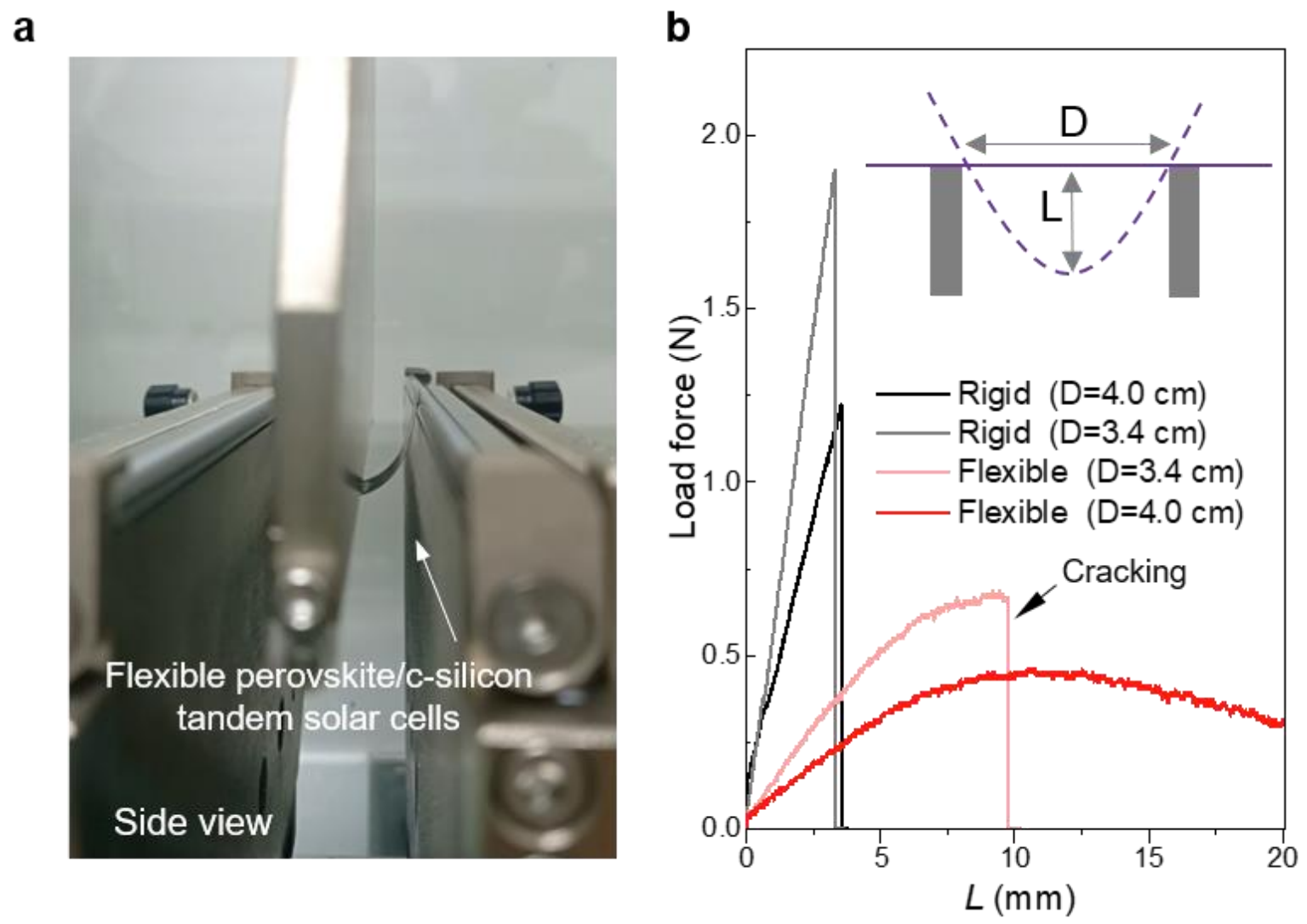

**Supplementary Fig. 18 | a,** Diagram of the Discovery dynamic mechanical analyzer (DMA) 850 for three-point bending test. **b,** Load-vertical displacement (F-L) curves of flexible and rigid devices, with load span of 3.4 and 4 cm, L represents the vertical displacement.

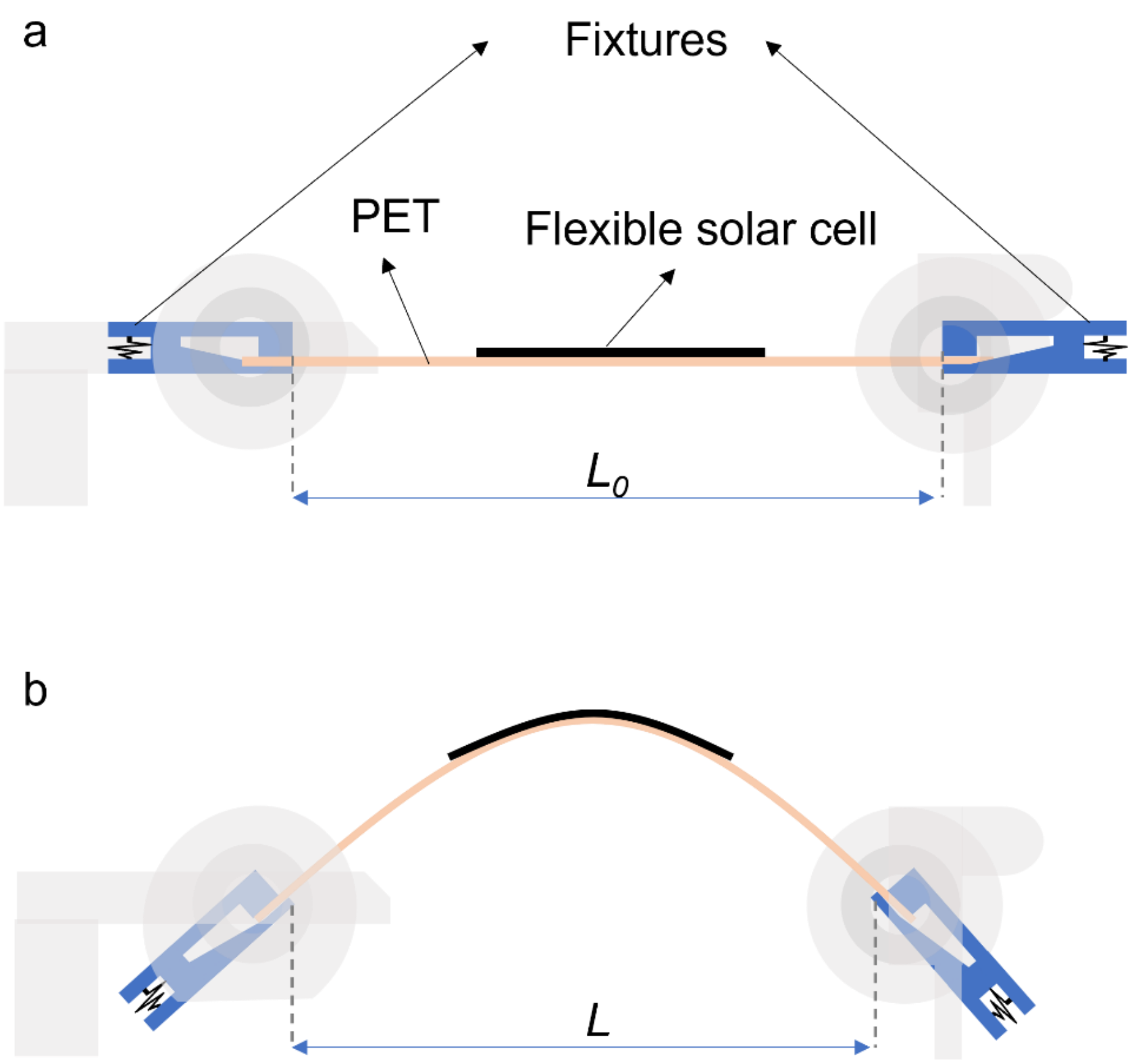

**Supplementary Fig. 19** | Schematics of the bending test. **a,** before bending and **b,** after bending. The bending radius (R) is calculated according following equation[25]:

$$R = \frac{L_0}{2\pi\sqrt{\frac{\Delta L}{L_0} - \frac{\pi^2 d^2}{12L_0^2}}}$$

where $L_0$ refers to the initial length of the sample before bending, $\Delta L$ refers to the shortened length after bending ($\Delta L = L_0 - L$), and $d$ refers to thickness of the substrate.

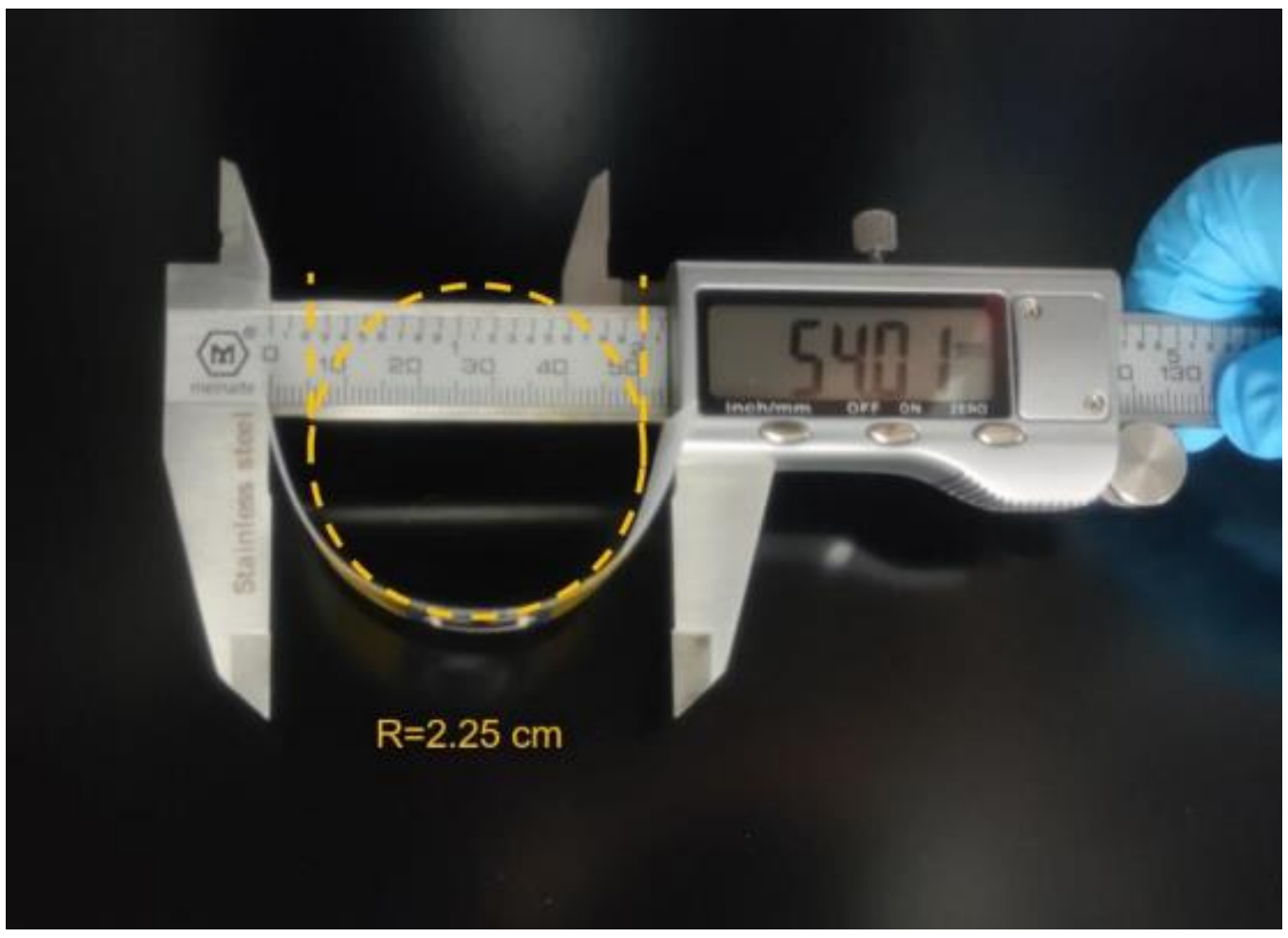

**Supplementary Fig. 20 |** The digital photo of flexible perovskite/c-Si monolithic tandem solar cell taken at the bending tests with a curvature radius of 2.25 cm.

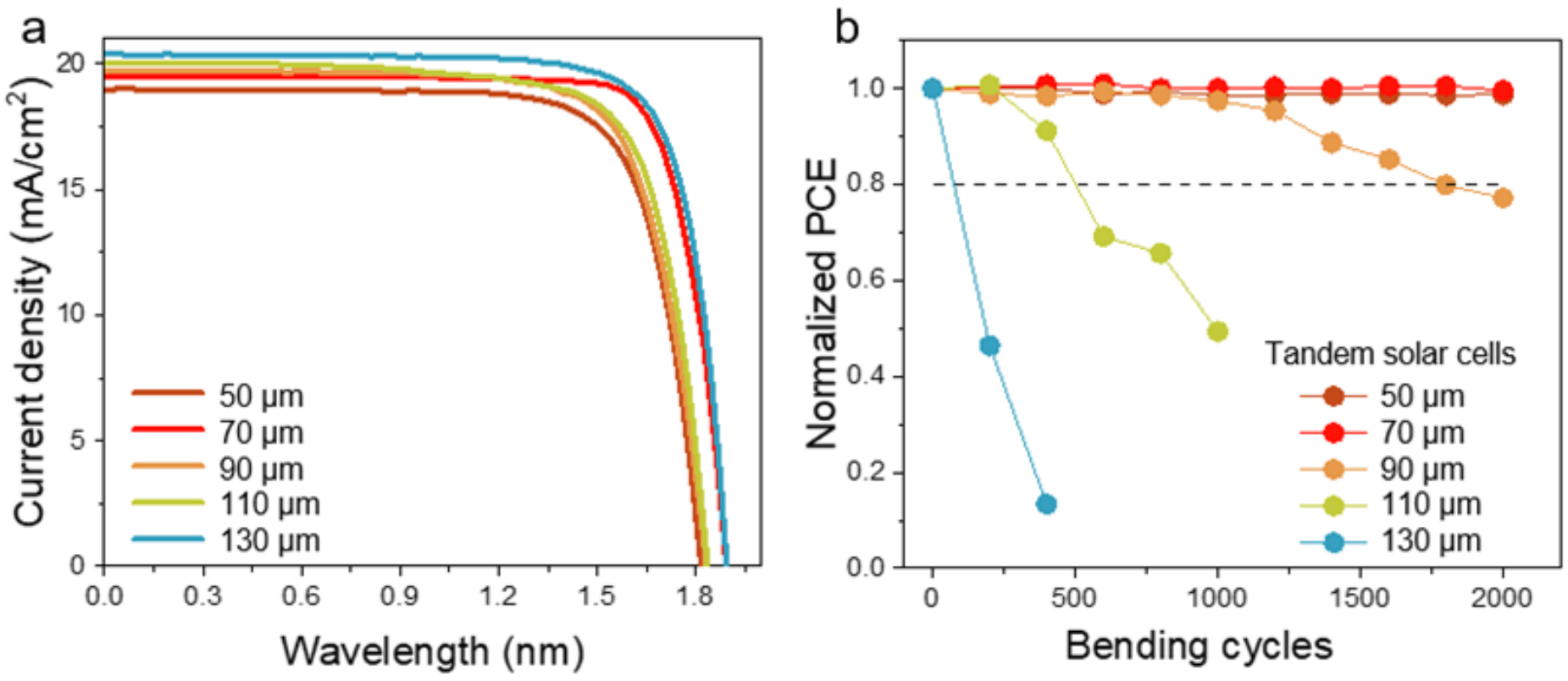

**Supplementary Fig. 21 | a,** PV performance of flexible perovskite/c-Si tandem solar cells with different silicon thickness and **b,** bending tests with different bending cycle numbers at the bending radius of 3.2 cm.

**Supplementary Table 1 |** PV parameters of flexile perovskite/silicon tandems with different silicon thickness.

| Thickness(μm) | $V_{OC}$ (V) | $J_{SC}$ (mA/cm$^2$) | FF (%) | PCE (%) |
|---|---|---|---|---|
| **50** | 1.814 | 18.98 | 76.19 | 26.25 |
| **70** | 1.890 | 19.52 | 80.90 | 29.83 |
| **90** | 1.829 | 19.94 | 75.44 | 27.20 |
| **110** | 1.833 | 20.03 | 75.16 | 27.62 |
| **130** | 1.891 | 20.38 | 78.79 | 30.36 |